\documentclass[1p,12pt,authoryear]{elsarticle}

\usepackage{amssymb}
\usepackage{amsthm}
\usepackage{lineno}
\usepackage{graphicx}
\usepackage[ruled,vlined]{algorithm2e}

\usepackage{soul,xcolor,color}
\usepackage{setspace}
\usepackage{amsmath}
\usepackage{threeparttable} %For adding notes below tables
\graphicspath{{./Figures/}}
\usepackage{xr}
\usepackage{url}

\begin{document}

\begin{frontmatter}
\title{Inference of microporosity phase properties in heterogeneous carbonate rock with data assimilation techniques}
\date{March 2025}

\author[1]{Zhenkai Bo}
\author[1]{Ahmed H. Elsheikh}
\author[1]{Hannah P. Menke}
\author[1]{Julien Maes}
\author[2]{Tom Bultreys}
\author[1]{Kamaljit Singh \corref{cor1}}
\affiliation[1]{organization={Institute of GeoEnergy Engineering, Heriot-Watt University},
            city={Edinburgh},
            postcode={EH14 4AS}, 
            country={United Kingdom}}

\affiliation[2]{organization={Centre for X-ray Tomography (UGCT),  Ghent University},
            city={Ghent}, 
            country={Belgium}}
\cortext[cor1]{Corresponding author.}
\begin{abstract}
Accurate digital rock modeling of carbonate rocks is limited by the difficulty in acquiring morphological information on small-scale pore structures. Defined as microporosity phases in computed tomography (micro-CT) images, these small-scale pore structures may provide crucial connectivity between resolved pores (macroporosity). However, some carbonate rocks are heterogeneous, and high-resolution scans are resource-intensive, impeding comprehensive sampling of microporosity phases. In this context, we propose the usage of the ensemble smoother multiple data assimilation (ESMDA) algorithm to infer the multiphase flow properties of microporosity phases from experimental observations for digital rock modeling. The algorithm's effectiveness and compatibility are validated through a case study on a set of mm-scale Estaillades drainage image data. The case study applies ESMDA to two capillary pressure models to infer the multiphase flow properties of microporosity phases. The capillary pressure curve and saturation map were used as observations to predict wetting phase saturation at six capillary pressure steps during iterative data assimilation. The ESMDA algorithm demonstrates improved performance with increasingly comprehensive observation data inputs, achieving better prediction than recently published alternative techniques. Additionally, ESMDA can assess the consistency between various forward physical models and experimental observations, serving as a diagnostic tool for future characterization. Given the diverse application conditions, we propose that ESMDA can be a general method in the characterization workflow of carbonate rocks. 

\end{abstract}

\begin{keyword}
Carbonate rocks; Digital rock physics; Microporosity phases; Inverse modeling.
\end{keyword}

\end{frontmatter}

%\linenumbers
\pagenumbering{arabic}

\section{Introduction}
Flow behaviors in subsurface porous materials, e.g., carbonate and sandstone rocks, are central to research in geoenergy technologies such as carbon capture and storage \citep{krevor2023subsurface}, underground hydrogen storage \citep{pan2021underground,jangda2023pore}, mineral recovery \citep{tang2023pore}, and geothermal systems \citep{pandey2015fracture}. Fluid flow in porous rocks largely controls these technologies, as flow regimes, defined as flow behaviors under the same dimensionless groups in scaling theory, play a fundamental role across scales despite multiple physical processes involved \citep{menke2023channeling,tang2024controlled}. Characterizing fluid flow behaviors in various flow regimes requires conducting experimental measurements while monitoring boundary conditions to determine the flow properties, such as capillary pressure and relative permeability \citep{alyafei2016experimental,lysyy_hydrogen_2022}. Conventional core-flooding experiments are resource-intensive and can irreversibly alter the pore structure by inducing chemical/mechanical processes, e.g., dissolution. Focusing on this, recent advances in imaging technologies enable pore ($10^{-6} m$) to core-scale ($10^{-2} m$) fluid flow simulations on X-ray micro-tomography (micro-CT) images to study and predict rock flow properties under diverse conditions at relatively manageable costs \citep{andra2013digital}. While these advancements have set the foundation of digital rock physics, significant challenges remain, especially for some rock types such as carbonate rocks, which have multi-scale pore-throat structures. 

Direct numerical simulation (DNS) and pore network modeling (PNM) are two numerical simulation methods that have been routinely used to build the digital twin of rock samples and provide predictions on important fluid dynamics \citep{maes2024dispersivity,blunt2013pore,foroughi2024incorporation}. As they both rely on the detailed pore-throat morphology information from 3D micro-CT images for accurate property predictions, the inherent balance between field of view (FoV) and image resolution of micro-CT imaging techniques means that neither of the two methods can effectively describe the fluid flow dynamics from one set of single-scale carbonate rock images. These images (typically resolution greater than $3\;\mu m$) exhibit three grayscale value regions: solid, void, and microporosity phases. Here, microporosity phases refer to those voxels that contain unresolved porosity structures that are smaller or similar to the image voxel size \citep{wang2022anchoring,menke2022using}. Thus, given the presence of microporosity phases in carbonate rock micro-CT images, the full connectivity of rock samples cannot be predicted via conventional DNS and PNM simulations. In this regard, multi-scale modeling and simulation techniques, e.g., Stokes-Brinkman (S-B) flow simulation \citep{menke2022using} and Darcy-type network elements \citep{bultreys2015multi,wang2023imaging}, are proposed to enable modeling flow behaviors in both resolved void and unresolved microporosity phases \citep{ruspini2021multiscale}. 

Two methods are often used to inform the microporosity phase properties during multi-scale simulations. One is using high-resolution images with a smaller FoV from the same core sample. Such images are first processed under DNS or PNM to get local properties, such as permeability and relative permeability. The corresponding predicted properties are then treated as representative values of all the other microporosity phases during whole-core multi-scale simulations \citep{ruspini2021multiscale}. The porosity map calculated by difference images between the dry scan and brine-saturated scan is another typical method to define the porosity of microporosity phases \citep{wang2022anchoring,foroughi2024incorporation}. In this scenario, other flow properties are defined based on the porosity values under assumptions or high-level experimental data, e.g., the whole core capillary pressure curve. 

The sample size restriction associated with high-resolution scanning can lead to inevitable destruction of the core sample and corresponding information loss during sampling (the drilling or cutting of smaller samples) \citep{menke2022using}. Moreover, it is still difficult to derive other petrophysical properties from the porosity map if the mineral type is not constant. Therefore, the current multi-scale simulation needs a more robust method to define the property models for each Darcy-type element or microporosity voxel. Inverse modeling using X-ray CT images of fluid saturation distribution has been used to parameterize heterogeneous capillary pressure characteristics of digital rock models \citep{jackson2018characterizing,wang2022anchoring,an2023inverse}. Given there are a great number of parameters to be determined in multi-scale digital rock models, a manual trial-and-error regression is often required \citep{foroughi2024incorporation}. Also, because of the limited sampling, there is still a lack of a validation method to reveal the uncertainty introduced by segmenting microporosity regions directly into several phases \citep{wang2022anchoring}.

In this context, data assimilation algorithms can potentially provide a viable solution. These algorithms utilize numerical forward modeling to update model parameters from observation data, providing reliable data integration and parameter update in the fields of reservoir engineering, hydrology, and numerical weather prediction \citep{houtekamer2005ensemble,jung2018ensemble,liu2008investigation}. Common data assimilation methods that have been applied in reservoir history matching are particle swarm optimization  \citep{mohamed2011history}, genetic algorithm \citep{ballester2007parallel}, and the most popular ensemble Kalman filter (EnKF) \citep{evensen1994sequential}. While EnKF processes observations sequentially as they become available, the Ensemble Smoother (ES) processes all historical data simultaneously in a single batch update, offering greater computational efficiency for reservoir history matching applications. Building on this batch approach, the Ensemble Smoother with Multiple Data Assimilation (ESMDA) further improves performance by iteratively assimilating the same dataset multiple times with inflated observation errors, reducing the ensemble collapse issues common in standard ES \citep{emerick2013ensemble}. As a variant of the ensemble Kalman filter \citep{emerick2013ensemble}, ESMDA is recognized for solving high-dimensional inverse problems while being capable of quantifying the inherent uncertainty \citep{zhou2022deep}. Given these advantages in handling high-dimensional inverse problems with uncertainty quantification, ESMDA is well-suited for the challenging task of inferring microporosity properties from multi-scale physical measurements in digital rock characterization. As such, we propose the usage of ESMDA algorithm as a general method to infer microporosity properties based on versatile physical measurements.

In this study, we use a set of mm-scale Estaillades core sample drainage images as a case study to demonstrate the efficacy and compatibility of ESMDA to various experimental measurements and data availability. The overall structure of this paper is as follows. Section \ref{sec:methodology} introduces the ESMDA algorithm and the image processing methods. Section \ref{sec:results} presents the validation of the results after ESMDA regression. Section \ref{sec:discussion} provides a summary of the main findings from the results and the computational cost of the ESMDA algorithm in this study. Lastly, section \ref{sec:conclusion} presents the concluding remarks.

\section{Methodology}\label{sec:methodology}
This section details the ESMDA methodology and demonstrates its implementation under different data availability and assumptions.

\subsection{Image sample description}
Many multi-scale simulation studies of carbonate rock samples use porosity-based segmentation to define the flow properties of microporosity regions \citep{menke2022using,foroughi2024incorporation}. Focusing on this, \cite{wang2022anchoring} scanned a series of saturation maps during a stepwise capillary-dominant core-flooding experiment (maximum capillary number $6.2\times10^{-8}$) on a mm-scale ($6.1 \times 6.1 \times 7 \; mm$) Estaillades rock sample for setting up a capillary pressure digital rock model. Since the capillary pressure during the core-flooding experiment is increased incrementally via a capillary plate, they are able to fit a Brooks-Corey (BC) model for each microporosity voxel. Using these voxel-wise BC models, microporosity regions can be segmented based on their threshold capillary pressure and porosity with the K-means cluster method, defined as capillary pressure-based segmentation. For each segmented microporosity phase, they use one set of averaged BC parameters of all the voxels within that phase to represent the capillary flow behaviors \citep{wang2022anchoring}. Capillary pressure-based and porosity-based multi-scale PNM models are set up accordingly. Lastly, they compare the prediction error between the two multi-scale PNM results with experimental saturation images. The capillary pressure-based segmentation shows much smaller voxel absolute mean saturation error compared to the porosity-based segmentation.  

While their voxel-wise approach demonstrates improved accuracy, capillary shielding can cause the fitted BC model for a given microporosity voxel to reflect the behavior of neighboring regions rather than their own. Also, considering more experimental measurements to be included to inform the digital rock models in the future, there is a need for more efficient and comprehensive parameter regression methods. This case study uses the raw images from \cite{wang2022anchoring} to demonstrate how ESMDA can harness multi-dimensional data from experiments to infer the capillary pressure model of microporosity phases and provide physical insights through ensemble analysis. In the following sections, we present the method to process the original micro-CT images and the implementation details of ESMDA for the inference of the BC model of each microporosity phase.

\subsection{Image processing}
\begin{figure}[htp!]
  \centering
    \includegraphics[width=1.0\textwidth]{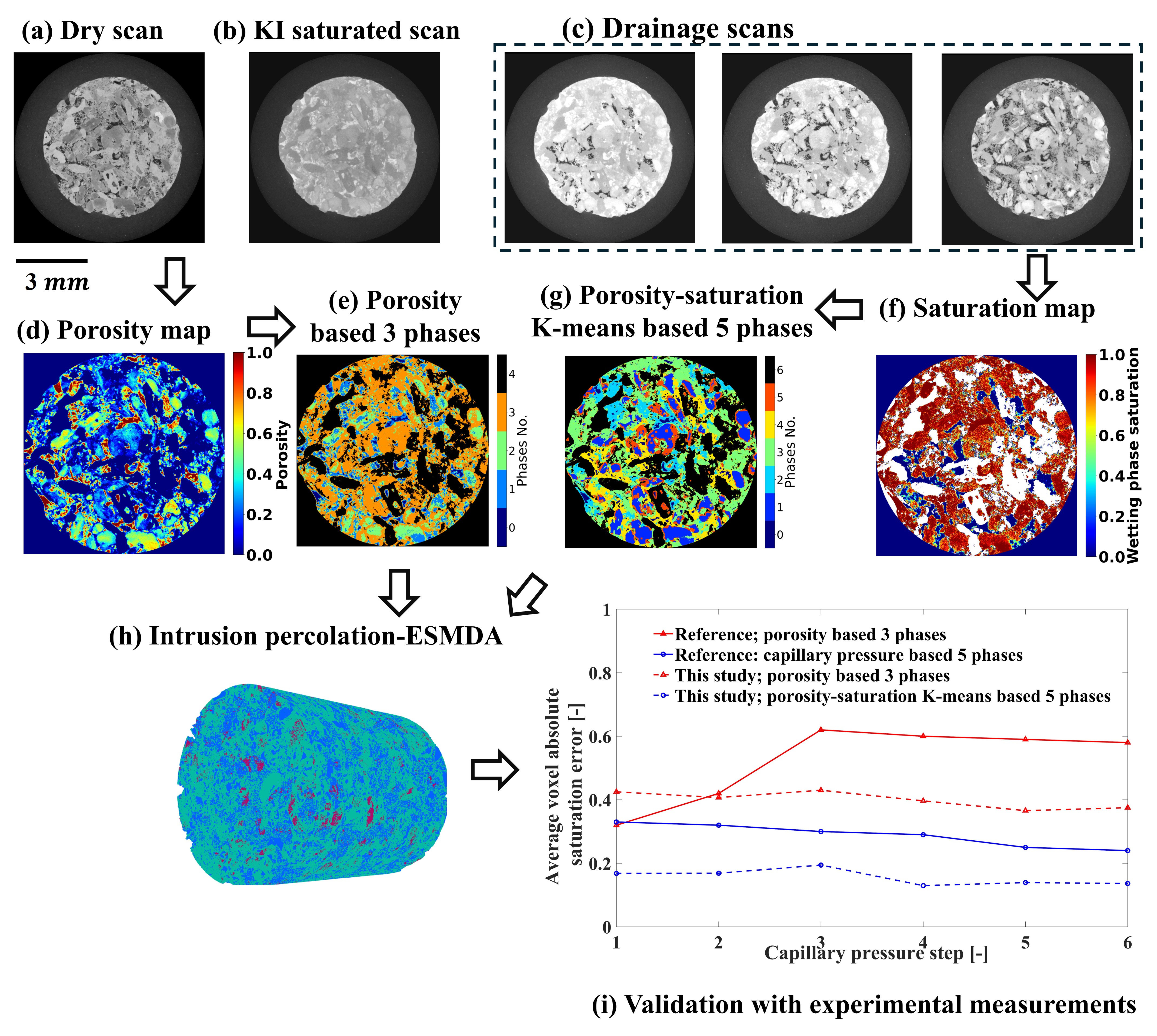}
    \caption{Schematic illustration of intrusion percolation-ESMDA workflow for capillary pressure model inference of prorosity based 3-phase and porosity-saturation based 5-phase segmentation models of Estaillades samples from \citep{wang2022anchoring}; (a-c) Raw images of dry scan, KI saturated wet scan, and drainage scans during stepwise drainage experiments under increasing capillary pressure; (d) Porosity map from difference images between dry scan and KI saturated wet scan; (e) Porosity based 3-phase segmentation model, phase 0 as resolved pores and phase 4 as solid phase; (f) Saturation map from difference images between dry scan and 6 drainage saturation scans; (g) Porosity-saturation K-means based 5 phases segmentation model, phase 0 as resolved pores and phase 6 as solid phase; (h) Intrusion percolation-ESMDA linearly updates the BC models for two segmentations models based on corresponding observations; and (i) Comparing simulated saturation map with experimental saturation map.}
    \label{fig:Case_study_Estaillades_graphi_Abstract}
\end{figure}

Figure \ref{fig:Case_study_Estaillades_graphi_Abstract} presents a schematic illustration of the workflow for capillary pressure model inference of two types of segmentation models. The raw images from \cite{wang2022anchoring} are first processed with the non-local mean filter developed by \cite{spurin2024python}, then scaled against dry scan images. After scaling, the difference images of the KI saturated scans and the drainage scans are used to calculate the porosity and saturation maps at each capillary pressure step, respectively. With the porosity map, the final whole core porosity is calculated as $0.259$, which is consistent with previous measurements in the literature ($0.247$ to $0.28$) \citep{alyafei2015sensitivity,bauer2012improving}. The full details of image processing can be found in \cite{wang2022anchoring} and \ref{App_section_image_processing}.

\subsection{Ensemble smoother with multiple data assimilation (ESMDA)}
In this study, the flow properties of each microporosity phase serve as input parameters for the multi-scale simulation on the digital rock model of carbonate rocks, and these parameters can be arranged into a vector $\mathbf{m}$ with dimension $N_m$. Before any experimental measurement, the probability distribution function of $\mathbf{m}$ is denoted as prior $\pi(\mathbf{m})$, which captures all the prior knowledge of the parameters $\mathbf{m}$ based on experience. The collected observations are stored in the vector $\mathbf{d}$. Given the observation error of $\mathbf{\epsilon}$ of the same dimension $N_d$, the experimental observation could be related to the forward multi-scale simulation model $\mathbf{g()}$ using the following equation: 

\begin{equation}\label{eq:ESMDA_Observation}
    \begin{aligned}
         \mathbf{d} = \mathbf{g(m)}+\mathbf{\epsilon}
    \end{aligned}
\end{equation}
To set up a proper digital rock model for carbonate rocks, our goal is to update the prior distribution of $\pi(\mathbf{m})$ via assimilating experimental measurement $\mathbf{d}$, to obtain the posterior distribution of multi-scale simulation parameter $\pi(\mathbf{m}\mid\mathbf{d})$. This is achieved through Bayes' rule as 

\begin{equation}\label{eq:Bayes_rule}
    \begin{aligned}
         \pi(\mathbf{m}\mid\mathbf{d}) = \frac{\pi(\mathbf{m})\pi(\mathbf{d}\mid\mathbf{m})}{\pi(\mathbf{d})}, \; \pi(\mathbf{d}) = \int \pi(\mathbf{m})\pi(\mathbf{d}\mid\mathbf{m})d\mathbf{m}
    \end{aligned}
\end{equation}
where $\pi(\mathbf{d}\mid\mathbf{m})$ is the likelihood function, and $\pi(\mathbf{d})$ is the evidence that serves as a normalizing constant \citep{tarantola2005inverse}.

To compute equation \ref{eq:Bayes_rule}, we use ESMDA \citep{emerick2013ensemble}, which is a variant of ensemble smoother (ES) or ensemble Kalman filter (EnKF). ES is initiated by drawing $N_e$ forecast (before assimilation) samples $\mathbf{M^{\textit{f}}}= [\mathbf{m}^{\textit{f}}_1,...\mathbf{m}^{\textit{f}}_{Ne}]$ from prior distribution $\pi(\mathbf{m})$, then linearly update them with 

\begin{equation}\label{eq:Ensemble_smoother}
    \begin{aligned}
         \mathbf{m}^a_j = \mathbf{m}^{\textit{f}}_j +\mathbf{C}^{\textit{f}}_{\mathbf{MD}}(\mathbf{C}^{\textit{f}}_{\mathbf{DD}}+\mathbf{C}_{\mathbf{D}})^{-1}[\mathbf{d}_{uc,j}-\mathbf{g}(\mathbf{m^\textit{f}_j})]
    \end{aligned}
\end{equation}
where $\mathbf{m}^a_j$ is the analysis ensembles conditioned on observation $\mathbf{d}$, $\mathbf{C}_D$ is the covariance matrix of observation error $\mathbf{\epsilon}$, $\mathbf{C}^{\textit{f}}_{\mathbf{DD}}$ is the auto-covariance of forward model predictions $\mathbf{D}^f=[\mathbf{m^\textit{f}_1},...,\mathbf{m^\textit{f}_{Ne}}]$, $\mathbf{g}(\mathbf{m^\textit{f}_j})$ is the forward model prediction with ensemble $j$, $\mathbf{C}^f_{MD}$ is the cross-covariance between $\mathbf{M}^f$ and $\mathbf{D}^f$, lastly $d_{uc,j}$ is the perturbed observation sampling from Gaussian distribution $\mathcal{N}(\mathbf{d}, \mathbf{C}_D)$.

On the basis of ES, ESMDA employs multiple iterations of ES with an inflated covariance matrix to damp parameter changes at the early iterations. For iteration index $i = 1,...N_a$, our ESMDA implementation is 

\begin{equation}\label{eq:ESMDA_iterations}
    \begin{aligned}
         \mathbf{m}^{i+1}_j = \mathbf{m}^{\textit{i}}_j +\mathbf{C}^{\textit{i}}_{\mathbf{MD}}(\mathbf{C}^{\textit{i}}_{\mathbf{DD}}+\alpha_i\mathbf{C}_{\mathbf{D}})^{-1}[\mathbf{d}^i_{uc,j}-\mathbf{g}(\mathbf{m^\textit{i}_j})]
    \end{aligned}
\end{equation}
where $\sum^{N_a}_{i=1}\alpha_i = 1$ to ensure consistency with ES, and $d^i_{uc,j}\sim \mathcal{N}(\mathbf{d},\alpha_i\mathbf{C_D})$. 

\subsection{Proposed microporosity phases property inference workflow}
\begin{figure}[htp!]
  \centering
    \includegraphics[width=1.0\textwidth]{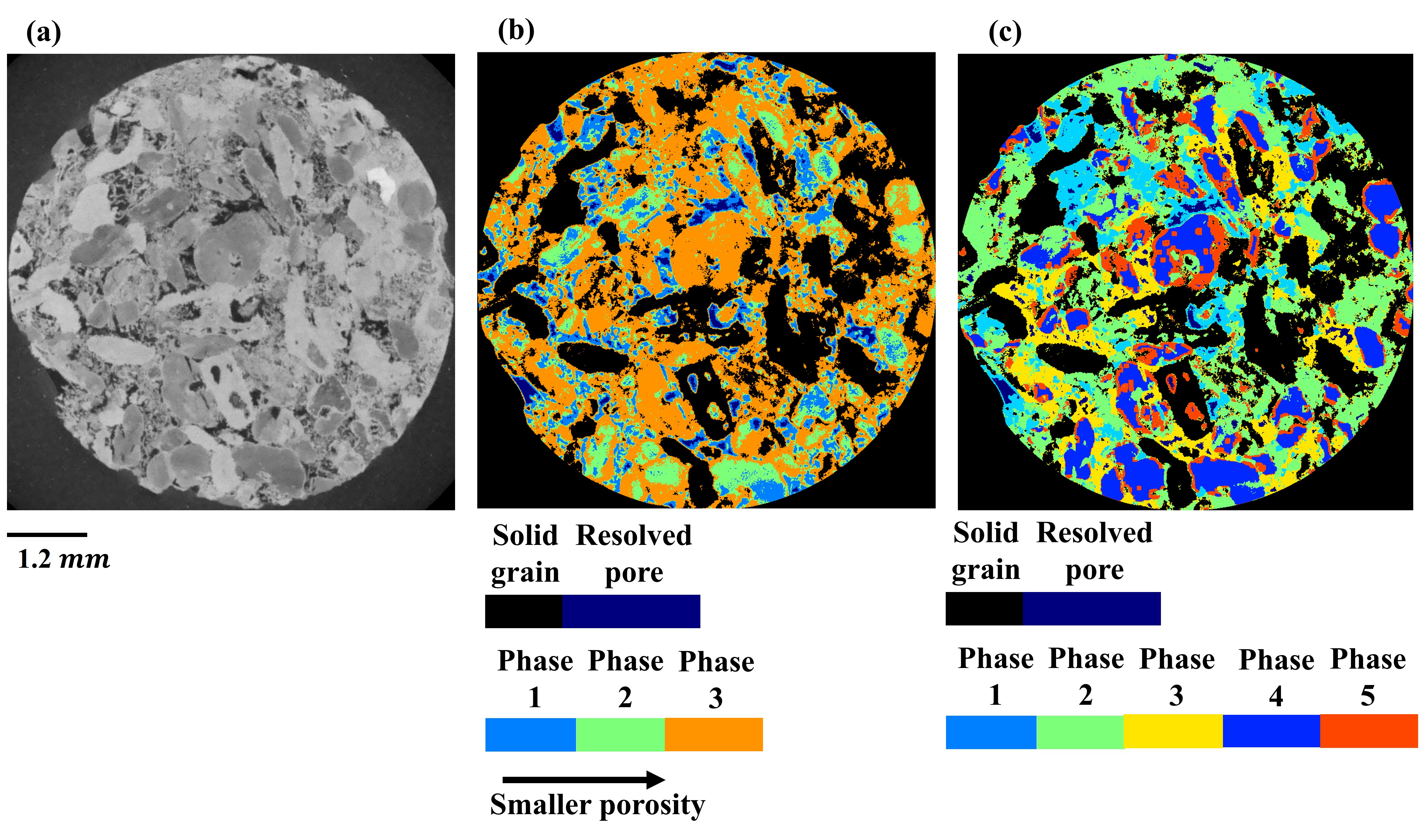}
    \caption{Cross section view of (a) dry scan raw image; (b) Porosity based 3-phase segmentation; (c) Porosity-saturation K-means based 5-phase segmentation model.}
    \label{fig:Case_study_Estaillades_segmentation_CrossSection}
\end{figure}

Porosity-based segmentation is usually performed based on voxel porosity distribution in previous studies \citep{foroughi2024incorporation,wang2022anchoring}. Here, we segment the microporosity region in the porosity map with two scenarios, as shown in Figure \ref{fig:Case_study_Estaillades_graphi_Abstract} d and f, which are porosity based 3-phase and porosity-saturation K-means based 5-phase segmentation models. The 3-phase model is designed for scenarios where experimental data are limited to porosity maps and whole-core capillary pressure curves that can be obtained without comprehensive step-wise core flooding experiments. This makes it more practical for most experimental conditions and easier to implement in future applications. Meanwhile, the 5-phase model represents the scenario where both porosity and saturation maps are available for ESMDA regression. Given the limited observation available for the 3-phase model, instead of setting thresholds from porosity distribution, we manually select porosity values of 0.6 and 0.4 as the thresholds for the 3-phase segmentation model. This is to test the robustness of ESMDA under non-ideal conditions, e.g., connectivity between resolved pores is unknown at each capillary pressure step in this study. For the 5-phase model, the spatial and statistical distributions of porosity and saturation at each capillary pressure step were analyzed using k-means clustering implemented through the \textit{KMeans} algorithm in the \textit{scikit-learn} package \citep{scikit-learn}. The final segmentation is shown in Figure \ref{fig:Case_study_Estaillades_segmentation_CrossSection}. 

Both the 3-phase and 5-phase models are fitted against observation data, but under different data availability settings. The observation data for the 3-phase model is limited as only the whole core capillary pressure curve is available. Meanwhile, the capillary pressure curve for each phase (derived from the saturation maps) is used as the observation data for the 5-phase model. Entry capillary pressure of phase $i$ $P_{e\_i}$ and constant $\lambda_i$ in the Brooks-Corey-type \citep{brooks1964hydrau} equation \ref{eq:BC_model} are inferred based on observation data to set up the capillary pressure model for each microporosity phase.

\begin{equation}\label{eq:BC_model}
    \begin{aligned}
         P_{c} =P_{e\_i}\bigg(\frac{1}{S_{w}}\bigg)^{{1}/{\lambda_i}}
    \end{aligned}
\end{equation}
where $i$ represents the number of phases. Accordingly, the intrusion-percolation simulation calculates the saturation of each microporosity phase based on equation \ref{eq:BC_model}. Note that every voxel is assumed to be connected from the first capillary pressure step, meaning all the resolved pore voxels' wetting phase saturation is calculated as 0, and the saturation of every microporosity phase voxel is calculated with equation \ref{eq:BC_model} at each capillary pressure step. This is defined as unconstrained cases in this study. The reason behind such a simulation strategy is twofold. One is the connectivity information from the core sample inlet end is lost as only the middle section images of the core sample are available, and this would be also similar to a real experimental scenario where there lack of saturation maps during drainage experiments. In addition, a forward physical model that can get saturation profile correct might not be computationally feasible to be implemented with ESMDA. Thus, we use such an unconstrained simulation strategy to showcase the capability of the ESMDA algorithm in typical carbonate rock characterization scenarios. Lastly, we use the following algorithm to implement the intrusion percolation-ESMDA operation, while the table \ref{Table Input_output_ESMDA} shows an overview of the algorithm input and output for both models.

% \begin{itemize}
% \item  Set the number of ESMDA iterations $N_a$ and corresponding $\alpha_i$. Generate initial ensemble $\lambda$ and $P_e$ values $\mathbf{m^{f}_{j}(j=1,...,N_e)$ from prior distribution $\mathbf{\pi(m)}$
% \item  Perform loop operation for $i = 1,...,N_a$:
% \begin{enumerate}
% \item Run forward intrusion percolation for each ensemble member $\mathbf{m}^{f}_{j}(j=1,...,N_e)$ to get the predicted saturation $\mathbf{g}(\mathbf{m}^{i}_{j})$.
% \item Perturb the observation $\mathbf{d}$ with inflated measurement noise $\alpha_i\mathbf{C_D}$, get $d^i_{uc,j}$. 
% \item Compute the cross variance matrix $\mathbf{C}_{MD}$ and auto-covariance matrix of predicted saturation $\mathbf{C}_{DD}$.
% \item Update the ensemble with the equation \ref{eq:ESMDA_iterations}, get $\mathbf{m}^{i+1}_{j}$ for the next iteration.
% \end{enumerate}
% \end{itemize}

\begin{algorithm}[H]
\caption{ESMDA Iterative Update}
Set number of ESMDA iterations $N_a$ and corresponding inflation coefficients $\alpha_i$\;
Generate initial ensemble $\mathbf{m}^{i=1}_{j}(j=1,\dots,N_e)$ of $\lambda$ and $P_e$ from prior distribution $\pi(\mathbf{m})$\;

\For{$i \gets 1$ \KwTo $N_a$}{
    \For{$j \gets 1$ \KwTo $N_e$}{
        Run forward intrusion percolation for ensemble member $\mathbf{m}^{i}_{j}$ to obtain predicted saturation $\mathbf{g}(\mathbf{m}^{i}_{j})$\;
        Perturb observations $\mathbf{d}$ with inflated noise $\alpha_i \mathbf{C_D}$ to obtain $d^i_{uc,j}$\;
    }
    Compute cross covariance matrix $\mathbf{C}^{i}_{MD}$ and auto-covariance of predicted saturation $\mathbf{C}^{i}_{DD}$\;
    Update ensemble using Eq.~\eqref{eq:ESMDA_iterations} to obtain $\mathbf{m}^{i+1}_{j}(j=1,\dots,N_e)$\;
}
\end{algorithm}

\begin{table}
\caption{Overview of input and output parameters of the ESMDA algorithm}
\label{Table Input_output_ESMDA}
\begin{center}
\begin{tabular}{c c c p{3.5 cm}}
\hline 
Models & Input $\mathbf{m_j^i}$ & Output $\mathbf{m_j^{i+1}}$ & Observation $\mathbf{d}$ \\
\hline
Estaillades 3-phase & $P_{e\_i}$ and $\lambda_i$ & $P_{e\_i+1}$ and $\lambda_{i+1}$& whole core capillary pressure curve \\
Estaillades 5-phase & $P_{e\_i}$ and $\lambda_i$ & $P_{e\_i+1}$ and $\lambda_{i+1}$  &  capillary pressure curve of each phase (from saturation map) \\
\hline
\end{tabular}
\end{center}
\end{table}

To quantify the saturation error at each step of the two models, we calculate the average voxel root mean square saturation error $\delta_{abs}$ as equation \ref{eq:Abs_root_mean_square_error}.

\begin{equation}\label{eq:Abs_root_mean_square_error}
    \begin{aligned}
         \delta_{abs} =\sqrt{\frac{\sum(S_{w\_exp}-S_{w\_sim})^2}{N}}
    \end{aligned}
\end{equation}
where $S_{w\_exp}$ and $S_{w\_sim}$ are the voxel wetting phase saturation of experimental observation and the voxels related to wetting phase saturation of percolation simulation results, respectively, and $N$ is the total number of non-solid voxels.

\section{Results}\label{sec:results}

\subsection{Saturation error}
\begin{figure}[htp!]
  \centering
    \includegraphics[width=1.0\textwidth]{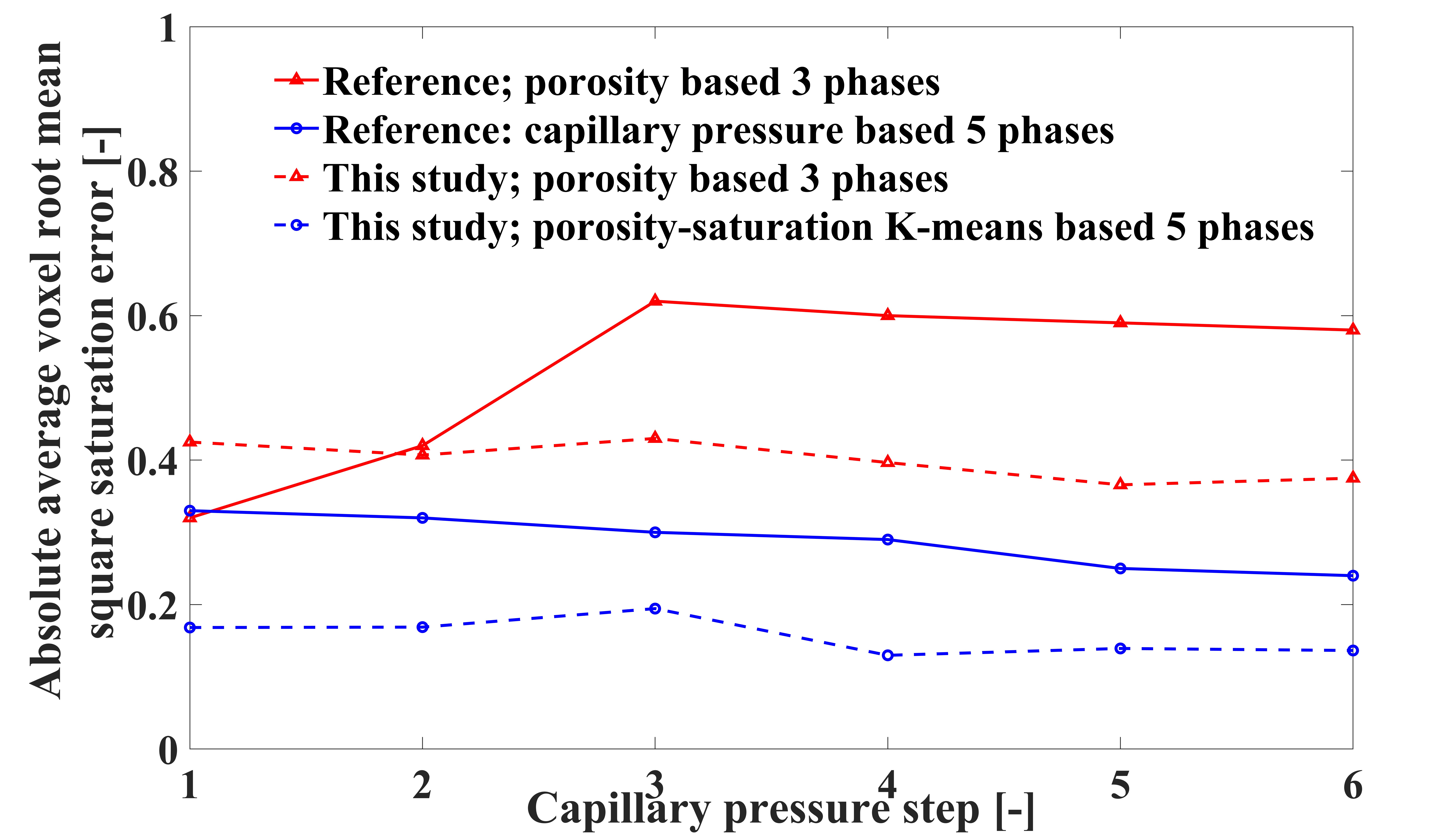}
    \caption{Average absolute voxel saturation error at each capillary pressure step of dashed red line: porosity based 3-phase model; dashed blue line: porosity-saturation based 5-phase model; solid red line: porosity based 3-phase model in \cite{wang2022anchoring}; solid blue line: threshold capillary pressure K-means based 5-phase model in \cite{wang2022anchoring}.}
    \label{fig:Voxel_saturation_abs_error_comparison}
\end{figure}

 Figure \ref{fig:Voxel_saturation_abs_error_comparison} shows the comparison between our saturation error $\delta_{abs}$ of 3-phase and 5-phase models with the porosity based 3-phase and threshold capillary pressure K-means based 5-phase model from \cite{wang2022anchoring}. Although we use the same raw data, our image filtering, regression, and percolation simulation methods differ in a few aspects. They use a multi-scale PNM to simulate the quasi-static drainage process and predict the saturation map. Whereas, we assume all the voxels are fully connected from the first capillary pressure step and calculate the saturation map based on the capillary pressure model of each microporosity phase directly. From this perspective, if ESMDA regression performs equal-effectively as the conventional voxel-wise regression presented in \cite{wang2022anchoring},  the $\delta_{abs}$ of our models should be several percent greater than those in \cite{wang2022anchoring}. This is true for the first capillary pressure step of our 3-phase model, shown in Figure \ref{fig:Voxel_saturation_abs_error_comparison}. However, other than this point, both our 3-phase and 5-phase models perform equally well or considerably better than their results. This further illustrates the superiority of ESMDA under various data availability conditions over the conventional regression method.

\subsection{Slice saturation comparison}
\begin{figure}[htp!]
  \centering
    \includegraphics[width=1.0\textwidth]{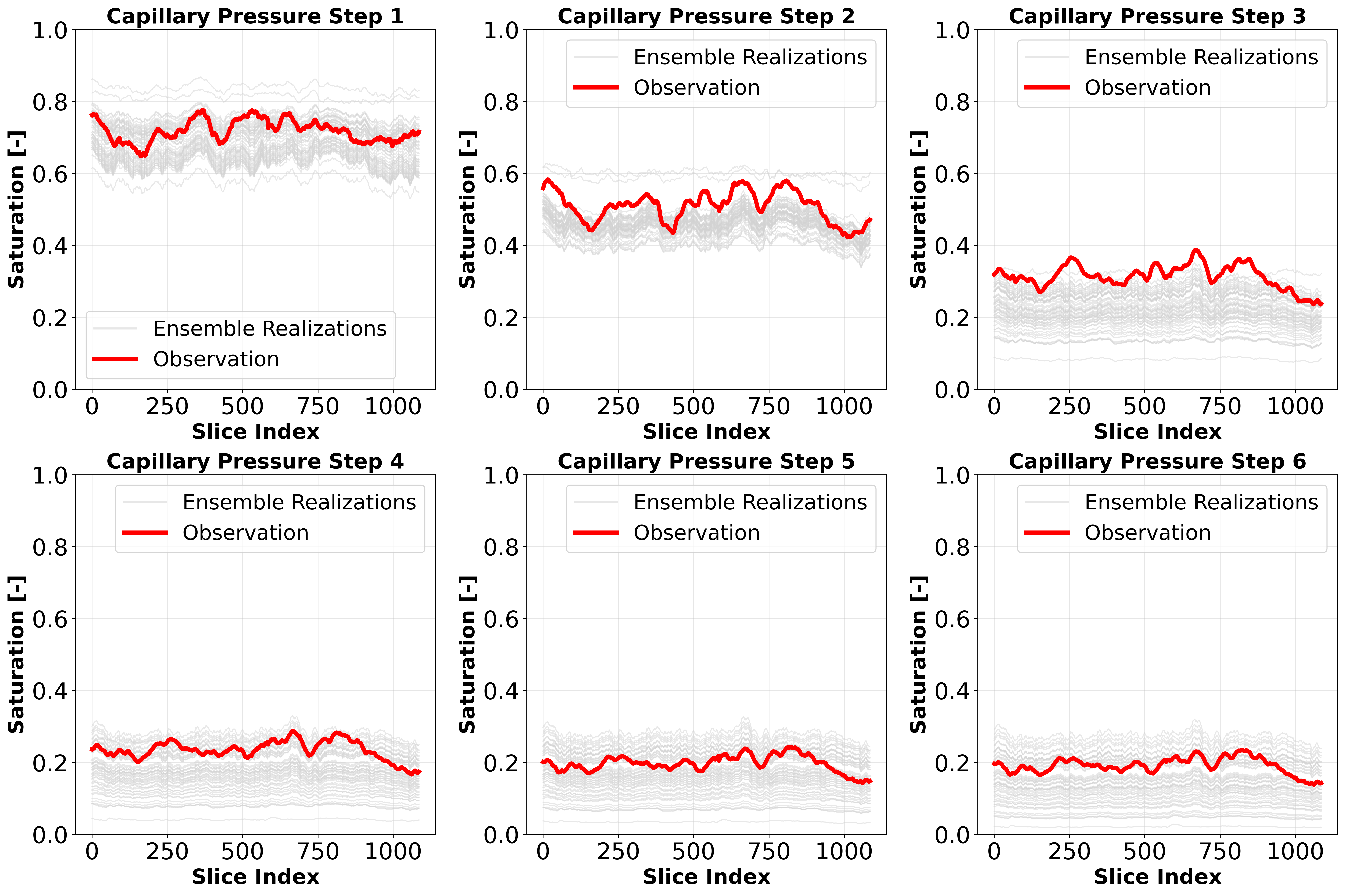}
    \caption{Slice saturation (along the length) comparison between the intrusion percolation simulation results of the porosity based 3-phase model, assuming all the voxels are connected, and the experimental images.}
    \label{fig:slice_saturation_comparison_3_phases_full_connection}
\end{figure}

\begin{figure}[htp!]
  \centering
    \includegraphics[width=1.0\textwidth]{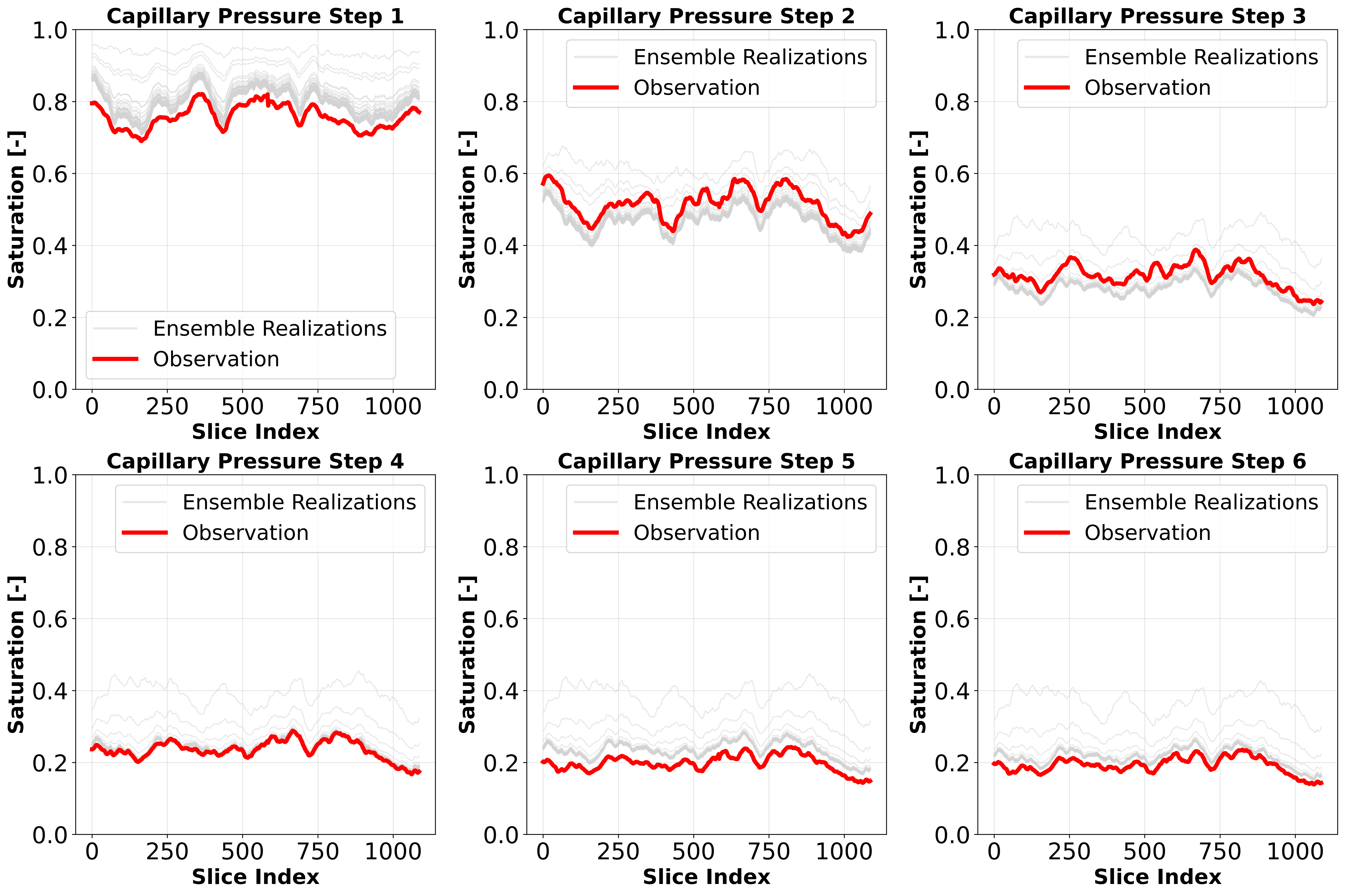}
    \caption{Slice saturation (along the length) comparison between the intrusion percolation simulation results of the porosity-saturation K-means based 5-phase model, assuming all the voxels are connected, and the experimental images.}
    \label{fig:slice_saturation_comparison_5_phases_full_connection}
\end{figure}

After ESMDA regression, we calculate the slice saturation of the best ensemble in terms of saturation error of both the 3-phase and 5-phase models and the experimental images with equation \ref{eq:slice_saturation_calculation}.  

\begin{equation}\label{eq:slice_saturation_calculation}
    \begin{aligned}
         S_{w\_i} =\frac{\sum S_{w\_ijk}\phi_{ijk}}{\sum \phi_{ijk}}
    \end{aligned}
\end{equation}
where $S_{w\_i}$ is the slice saturation of slice $i$, $ S_{w\_ijk}$ is the saturation of voxel at coordinate of $(i,j,k)$, $\phi_{ijk}$ is the porosity value of voxel at coordinate of $(i,j,k)$. Figure \ref{fig:slice_saturation_comparison_3_phases_full_connection} and \ref{fig:slice_saturation_comparison_5_phases_full_connection} show the comparison of intrusion percolation simulation (drainage process) results of the porosity based 3-phase model and the porosity-saturation K-means based 5-phase model with experimental image results, respectively.

The 3-phase model can only present the general trend of the first three capillary pressure steps. Specifically, the updated ensembles of 3-phase model (light gray curves) of the first three capillary pressure steps scatter around the observation (red curve) and show much more variability than the later 3 steps. Only some local trends, e.g., a decreasing trend between 300 and 500 slices at steps 1 and 2, and between 600 and 750 slices at steps 2 and 3, are captured. The reason for this failure to match such saturation variability is twofold. One is that, as discussed in previous sections, the experimental images are the middle section of the core sample, so the connectivity of the resolved pores is unknown, which leads us to assume all the resolved and microporosity voxels are connected during our intrusion percolation simulation. The other is that porosity-based segmentation does not consider the entry capillary pressure distribution of microporosity phases, leaving both the high and low saturation regions segmented into the same phase, see figure \ref{fig:Case_study_Estaillades_graphi_Abstract} (e) and (f). These factors smear the differences between microporosity phases having different entry capillary pressures, resulting in a failure to capture local extreme saturation values.  

\begin{figure}[htp!]
  \centering
    \includegraphics[width=1.0\textwidth]{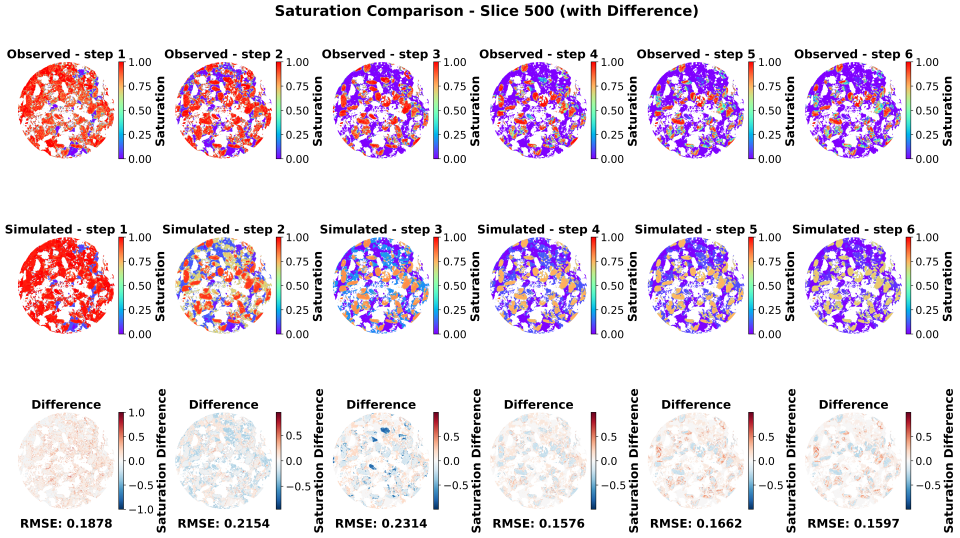}
    \caption{Cross-section saturation distribution comparison between the experimental observations (top row); the intrusion percolation simulation results of the porosity-saturation K-means based 5-phase model (middle row), assuming all the voxels are connected; and the difference map (bottom row) with root mean square error (RMSE).}
    \label{fig:saturation_map_comparison_5_phases_full_connection}
\end{figure}

In contrast, the 5-phase model successfully capture the general trend of all six capillary pressure steps and local extreme values of the latter three steps, such as the saturation minima around slice 750 at steps 4 to 6 in Figure \ref{fig:slice_saturation_comparison_5_phases_full_connection}. Given that the resolved pores are not fully invaded during the first 3 capillary pressure steps of experiments, the small discrepancy between the simulation results and the experimental observations is as expected under the full connectivity assumption. After these early steps most of the resolved pores are invaded and the detailed saturation profiles are fully captured, especially at steps 4 and 6, as shown in Figure \ref{fig:slice_saturation_comparison_5_phases_full_connection}. Additionally, we further examine the cross-section saturation distribution between the 5-phase model simulation results and the experimental images. Figure \ref{fig:saturation_map_comparison_5_phases_full_connection} gives the saturation map comparison where the model accurately captures the relative contrast between high and low saturation regions, with phase-wise saturation errors predominantly within 0.2. The primary source of this phase-wise saturation error, similar to the 3-phase model, is the large spread in saturation values at each porosity, as shown in Figure \ref{app_fig: Casestudy_Estaillades_Porosity-saturation_distribution_heatmap}. For any given porosity value, the saturation spans nearly the full range from 0 to 1, rather than being concentrated to a narrow range. Figure \ref{app_fig:Phases_segmentation_smearing_constrast} shows a microporosity phase in the 5-phase model that includes two distinct saturation clusters during experiments. As such, such saturation contrast within each phase would be averaged during regression, resulting in mild error presented in Figure \ref{fig:saturation_map_comparison_5_phases_full_connection}.

\section{Discussion} \label{sec:discussion}
\begin{figure}[htp!]
  \centering
    \includegraphics[width=1.0\textwidth]{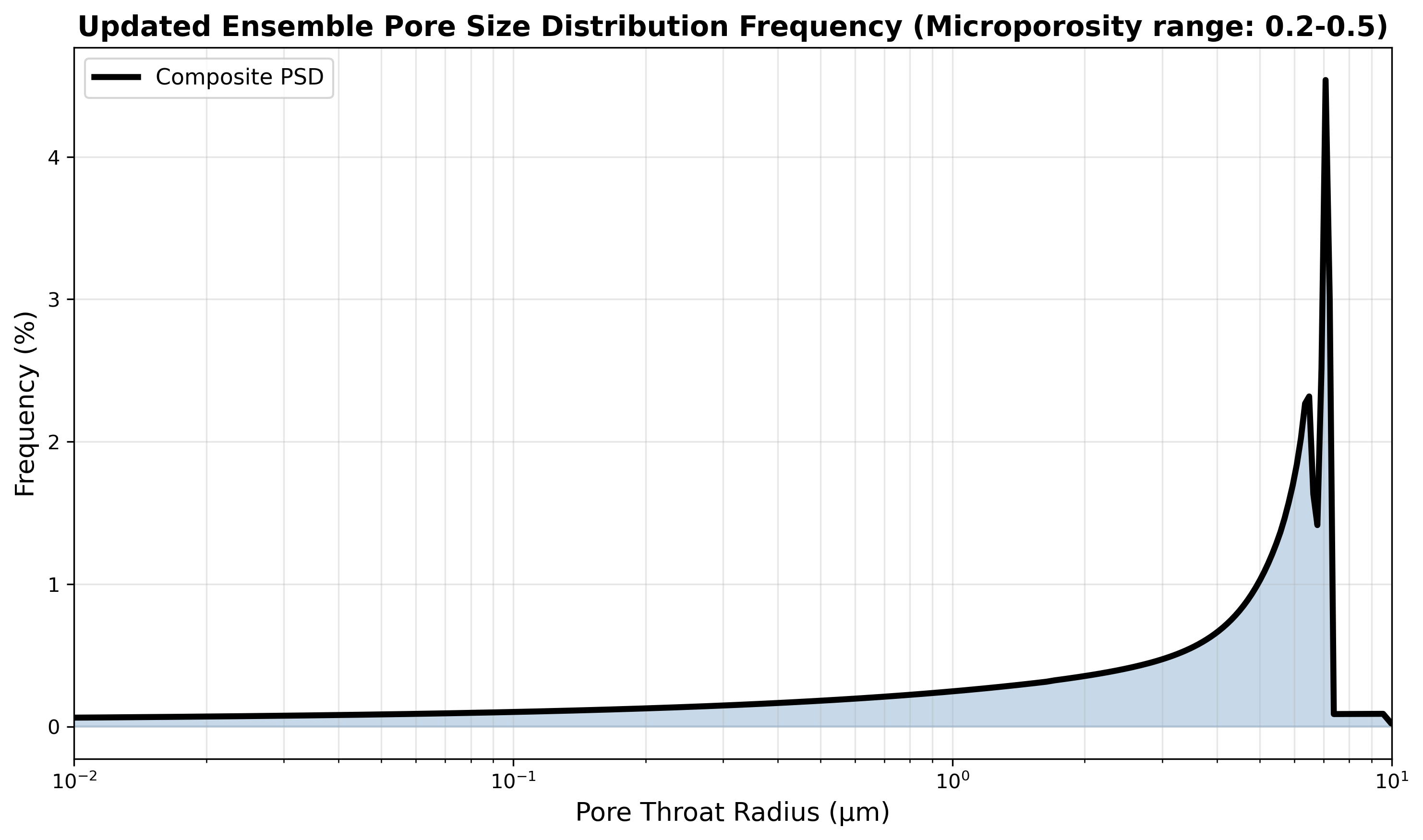}
    \caption{Pore size distribution of updated ensembles of unconstrained 5-phase model with Brooks-Corey and Young-Laplace equations (see \ref{app_sec: PSD}).}
    \label{fig:ESMDA_pore_size_distribution_5_phases_NoConnectivity}
\end{figure}

Due to the multi-scale and multi-physics nature of porous geological rock sample characterization, experimental or computational observations from various origins are used to build a digital twin of rock samples that reproduces the physical responses observed in measurements \citep{wang2022anchoring,foroughi2024incorporation,an2023inverse}. Such a process would not be straightforward and needs iterative implementation of a 'measurement-regression-validation' procedure. From this perspective, our results systematically demonstrate the good compatibility of ESMDA with this type of workflow. For example, the ESMDA algorithm can capture the information from limited observations to provide a general description of the core sample structure, as in Figure \ref{fig:slice_saturation_comparison_3_phases_full_connection}. As more measurement results (observations) become available, the predictive digital model established through the ESMDA algorithm keeps improving, see Figure \ref{app_fig:Slice_saturation_comparison_porosity_3_phases_model_connectivity_known}, where connectivity information is introduced on top of the whole core capillary pressure curve (connectivity constrained 3-phase model). Moreover, the resulting distribution of updated ensembles is physically realistic. Figure \ref{fig:ESMDA_pore_size_distribution_5_phases_NoConnectivity} shows the pore size distribution derived from Brooks-Corey and Young-Laplace equations (see \ref{app_sec: PSD}) with updated ensembles of unconstrained 5-phase model. The distribution shows two peaks in the 1-10 $\mu m$ range: one around 6 $\mu m$ and another around 7 $\mu m$, which is in a similar pattern to the pore size distribution of Estaillades carbonates from \cite{tanino2012capillary} except that their distribution includes an additional peak around 200 nm. The absence of this small pore size peak in our model is consistent with the systematic divergence observed in the low porosity range ($<0.4$) when comparing average wetting phase saturation distribution across porosity levels between experimental measurements and updated ensembles of both 3-phase and 5-phase models, as shown in Figure \ref{app_fig:Casestudy_Estaillades_AvgsSat-por_Distribution_3phases} and \ref{app_fig:Casestudy_Estaillades_AvgsSat-por_Distribution_5phases}. This divergence suggests that the model's inability to capture the finest pore structures (corresponding to low porosity regions) may explain both the missing peak and the saturation-porosity discrepancies. From an experimental point of view, part of the reason can be that, at high capillary pressure, the drainage process becomes very slow and the X-ray CT images did not capture the status that all the microporosity phases are invaded. 

\begin{figure}[htp!]
  \centering
    \includegraphics[width=1.0\textwidth]{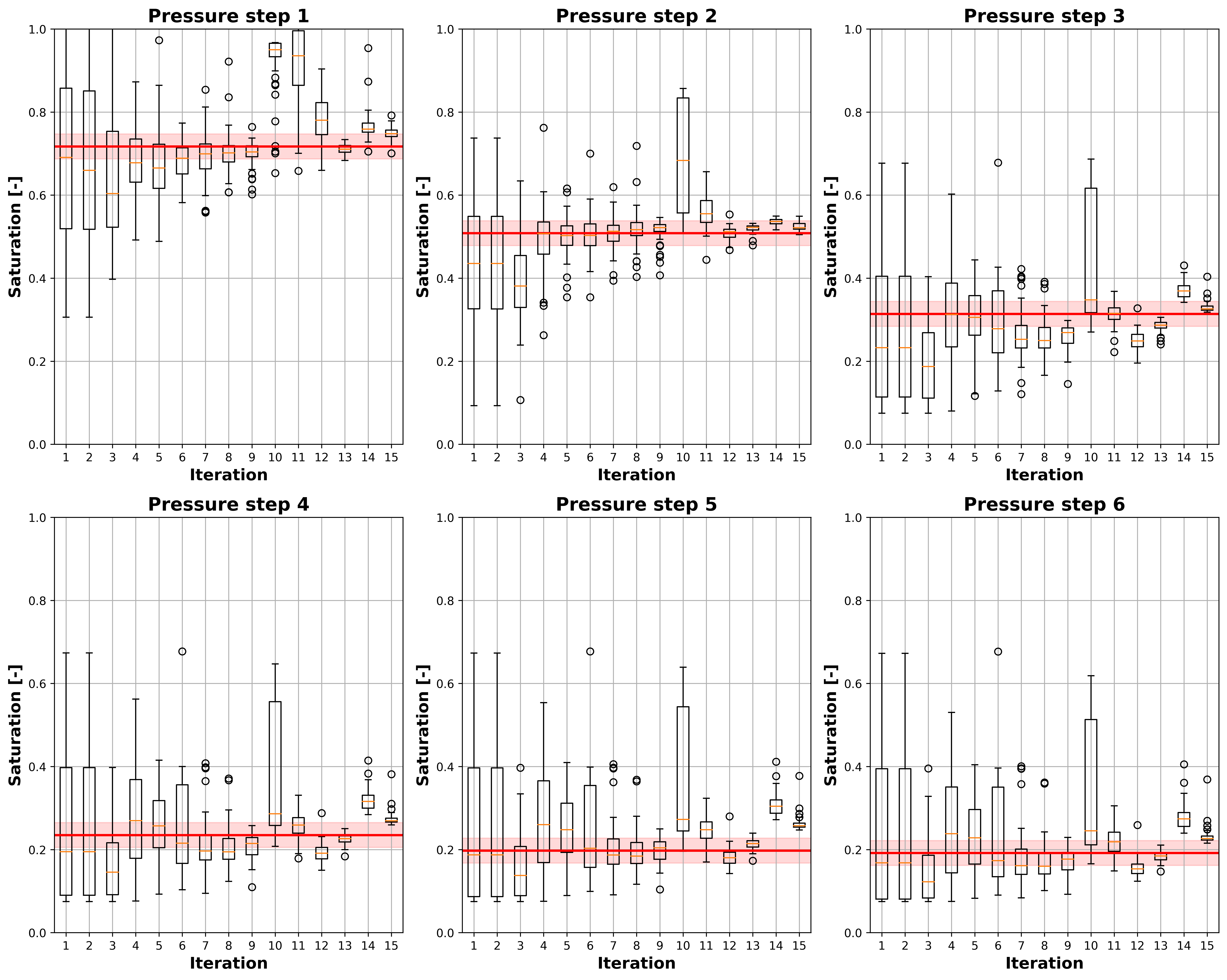}
    \caption{Box plot of updated ensembles during ESMDA regression of unconstrained porosity based 3-phase model assuming all the voxels are connected; boxes represent the $75^{th}$ to $25^{th}$ percentiles predicted whole core saturation of updated ensembles from the forward model and dots represent outliers.}
    \label{fig:ESMDA_regression_boxplot_3_phases_NoConnectivity}
\end{figure}

\begin{figure}[htp!]
  \centering
    \includegraphics[width=1.0\textwidth]{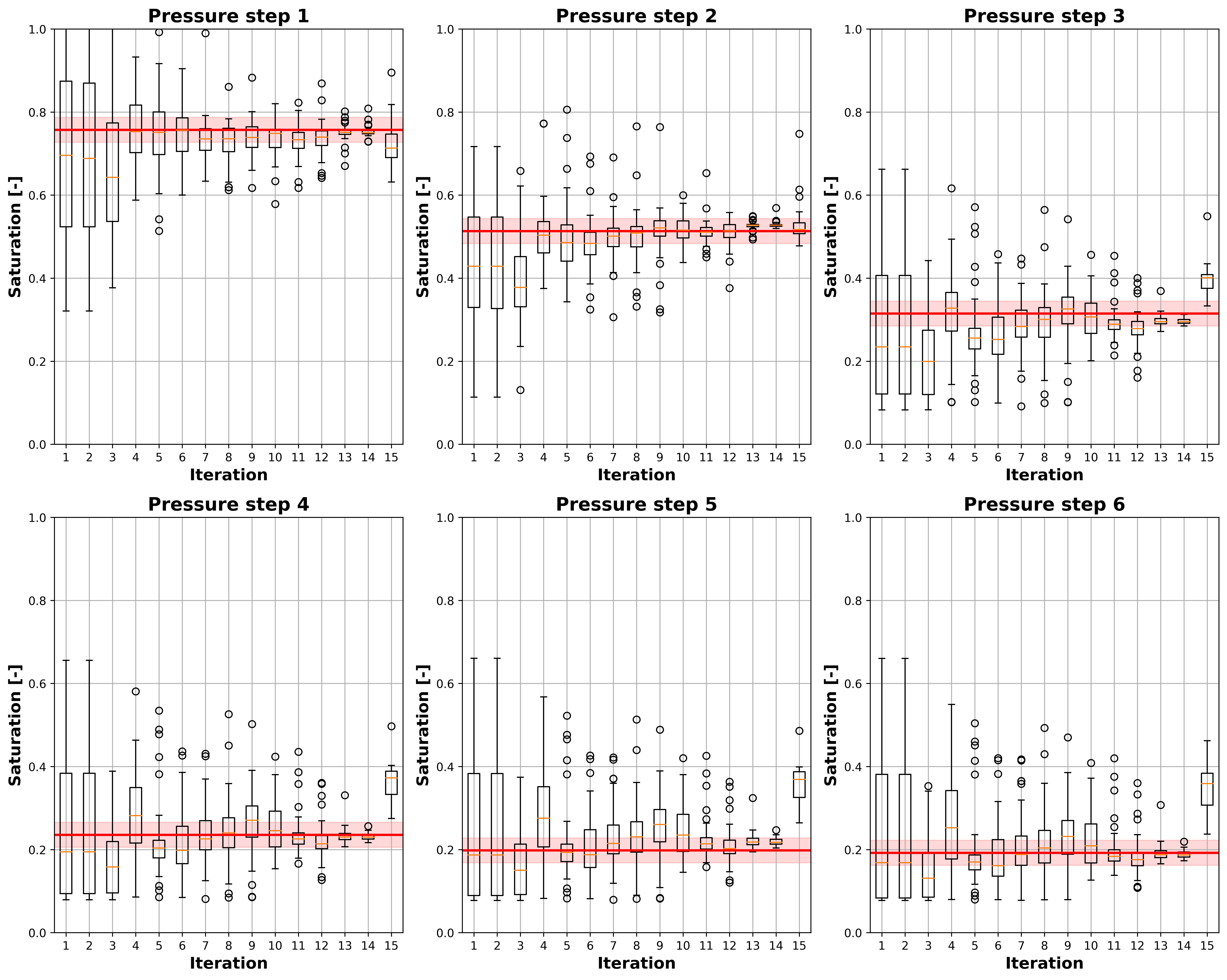}
    \caption{Box plot of updated ensembles during ESMDA regression of constrained porosity based 3-phase model assuming connectivity information is known; boxes represent the $75^{th}$ to $25^{th}$ percentiles predicted whole core saturation of updated ensembles from the forward model and dots represent outliers.}
    \label{fig:ESMDA_regression_boxplot_3_phases_Connectivity}
\end{figure}

\begin{figure}[htp!]
  \centering
    \includegraphics[width=1.0\textwidth]{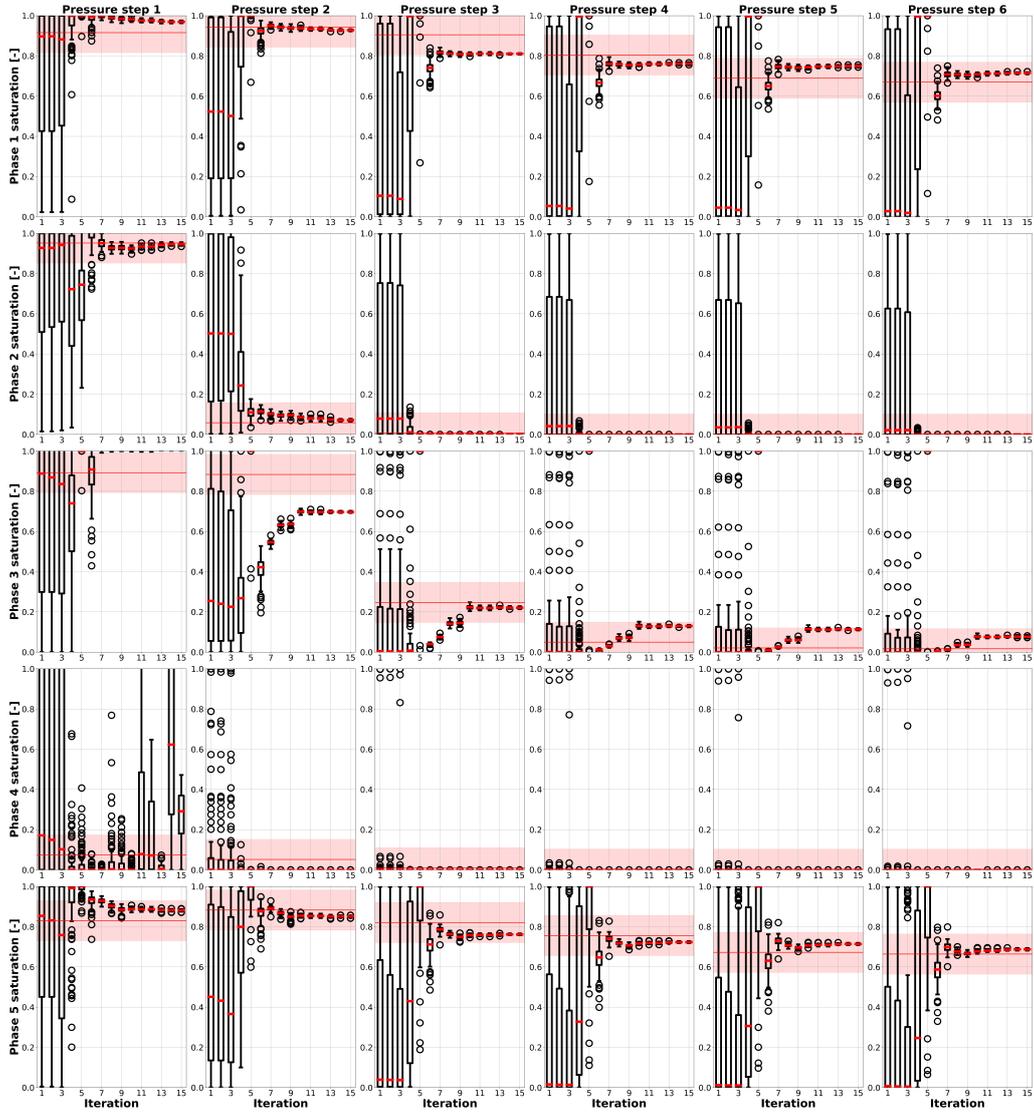}
    \caption{Box plot of updated ensembles during ESMDA regression of unconstrained porosity-saturation k-means based 5-phase model assuming all the voxels are connected; boxes represent the $75^{th}$ to $25^{th}$ percentiles predicted phase saturation of updated ensembles from the forward model and dots represent outliers.}
    \label{fig:ESMDA_regression_boxplot_5_phases_NoConnectivity}
\end{figure}

The uncertainty estimation by ESMDA on each parameter of the model and corresponding regression processes also reveals the consistency of the forward model to experimental observations, facilitating decision-making for more comprehensive future characterization. \cite{wang2022anchoring} raises the research question that whether the capillary shielding is the reason why the porosity-based 3-phase model performs much worse than the capillary pressure-based models. To examine this hypothesis, we compare the regression process and saturation predictions of connectivity constrained 3-phase model against the unconstrained 3-phase and 5-phase models. Here, the constrained model means the wetting phase saturation of uninvaded microporosity voxels during the experiment will be set as 1 and not be input into the ESMDA regression process. As such, the high wetting phase saturation of these microporosity voxels at early capillary pressure steps will not affect the regression process of other microporosity phases, preventing capillary shielding from incorrectly assigning the same capillary pressure parameters to the uninvaded microporosity voxels as their invaded neighbors during regression.  Figure \ref{fig:ESMDA_regression_boxplot_3_phases_NoConnectivity}, \ref{fig:ESMDA_regression_boxplot_3_phases_Connectivity}, and \ref{fig:ESMDA_regression_boxplot_5_phases_NoConnectivity} show that the regression processes of the unconstrained 5-phase model are more stable than both 3-phase models, regardless of the connectivity assumption. The three box plots reveal that the unconstrained 5-phase model exhibits more consistent convergence with reduced ensemble spread and fewer outliers during the last few iterations, while both 3-phase model show greater variability in parameter updates, with more pronounced fluctuations in the ensemble distributions and occasional divergent realizations during the regression process. Further, the slice saturation profile of the constrained 3-phase model shows less discrepancy between simulation and experimental observations compared to that of the unconstrained 3-phase model, as well as less ensemble uncertainty. However, its discrepancy and uncertainty are still much greater than those of the constrained 5-phase model, shown in Figure \ref{app_fig:Slice_saturation_comparison_porosity_3_phases_model_connectivity_known}. This greater discrepancy and uncertainty indicate weaker consistency between the 3-phase model and experimental observations, even when connectivity information is known (considering capillary shielding effects). Hence, while capillary shielding may contribute to the poor performance of the 3-phase model, it may not be the most important factor. Better segmentation that considers the spatial distribution of porosity and saturation history would also help resolve the problem. These results illustrate the diagnostic capability of ESMDA in revealing the consistency between the forward physical model and the experimental observations. To the best of our knowledge, there has been no technique that can achieve multi-parameter regression while estimating uncertainty in the field of porous rock characterization, similar to ESMDA presented in this study.

The computational cost of ESMDA is affordable. Table \ref{Table Run_time_ESMDA} provides an overview of the total run time of ESMDA regression for the two models presented in this study with an Intel(R) Xeon(R) Gold 6430 64-core CPU. Note that the run time refers to the execution time of the $N_a$ ESMDA iterations, where the data processing time, e.g., image processing, is omitted here. $N_a$ in Table \ref{Table Run_time_ESMDA} is the number of iterations that are needed to reach the first reasonable match. The regression processes in Figure \ref{fig:ESMDA_regression_boxplot_3_phases_NoConnectivity}, \ref{fig:ESMDA_regression_boxplot_3_phases_Connectivity}, and \ref{fig:ESMDA_regression_boxplot_5_phases_NoConnectivity} are extended to show the stability.

\begin{table}
\caption{Overview of total run time of ESMDA iterations of two models in this study}
\label{Table Run_time_ESMDA}
\begin{center}
\begin{tabular}{c c c c}
\hline 
Models & Ensemble size $N_e$ & Number of iterations $N_a$ & Total run time [s]\\
\hline
Estaillades 3-phase & 50 & 9 & 10.86 \\
Estaillades 5-phase & 100 & 9  &  8.06 \\
\hline
\end{tabular}
\end{center}
\end{table}

The major source of voxel saturation error after ESMDA in this study comes from the phase segmentation and underlying assumptions, instead of the regression method. As shown in Figure \ref{app_fig: Casestudy_Estaillades_Porosity-saturation_distribution_heatmap}, the distribution profile of porosity-saturation from experiments is highly dispersed, increasing the number of phases would therefore improve voxel-wise accuracy. This is consistent with the observation in the scatter plots of simulation voxel saturation against the experimental observation of 3-phase (Figure \ref{app_fig:Casestudy_Estaillades_saturation_comparison_scatter_heatmap_3phases}) and 5-phase models (Figure \ref{app_fig:Casestudy_Estaillades_saturation_comparison_scatter_heatmap_5phases}). The 3-phase model can not effectively segment the saturation distributions, as many capillary pressure steps show horizontal clustering in the scatter plots, where individual simulation saturation values map to broad ranges of experimental saturation observations. This demonstrates the model's inability to capture the wide saturation distribution presented in the experimental data. In contrast, 5-phase model improves the clustering behavior with data points more tightly distributed along the 1:1 line and correspondingly greater correlation coefficient. These observations explain the reason for the high voxel saturation error of both the porosity based 3-phase models in our study and \cite{wang2022anchoring}. Moreover, given the superiority of K-means based over porosity based segmentation methods on handling morphological data such as microporosity clusters in heterogeneous carbonates in this study, we believe machine learning would be a routinely used method in future digital rock modeling. From this perspective, ESMDA is well-suited for the workflow as the great compatibility of it to different machine learning algorithms.

\section{Conclusions}\label{sec:conclusion}
In this study, we propose the ensemble smoother with multiple data assimilation (ESMDA) algorithm as a method for inference of carbonate rock microporosity phase properties from experimental observations. A case study on an open source Estaillades drainage images is presented to demonstrate its efficacy, and we reach the following conclusions.

(1) ESMDA, as a multiple parameter regression method, is able to capture the underlying structure of carbonate rock microporosity phases from various data sources. With limited data, it can facilitate predicting a general trend of slice saturation distribution of each microporosity phase during the drainage core-flooding experiment. As the data availability improves, the ESMDA algorithm can deliver efficient regressions that are better than manual voxel-wise regression in terms of average voxel absolute saturation error.

(2) The uncertainty assessment capability of ESMDA on model parameter and ensemble regression process would inform the consistency of the forward physical model and experimental observations, facilitating more comprehensive future characterization.

%(3) The implementation of ESMDA in programming languages such as Python gives it great compatibility with machine learning. Consequently, many of the computationally dense and high-dimensional problems associated with carbonate rock sample micro-CT images, e.g., image processing, can be addressed through coupling machine learning with ESMDA. 

(3) Given the versatility and reasonable computational cost, the ESMDA will be suitable for building up digital models of heterogeneous carbonate rocks from multi-dimensional experimental data. 

\appendix
\section{Estaillades core sample image processing} \label{App_section_image_processing}
\begin{figure}[htp!]
  \centering
    \includegraphics[width=1.0\textwidth]{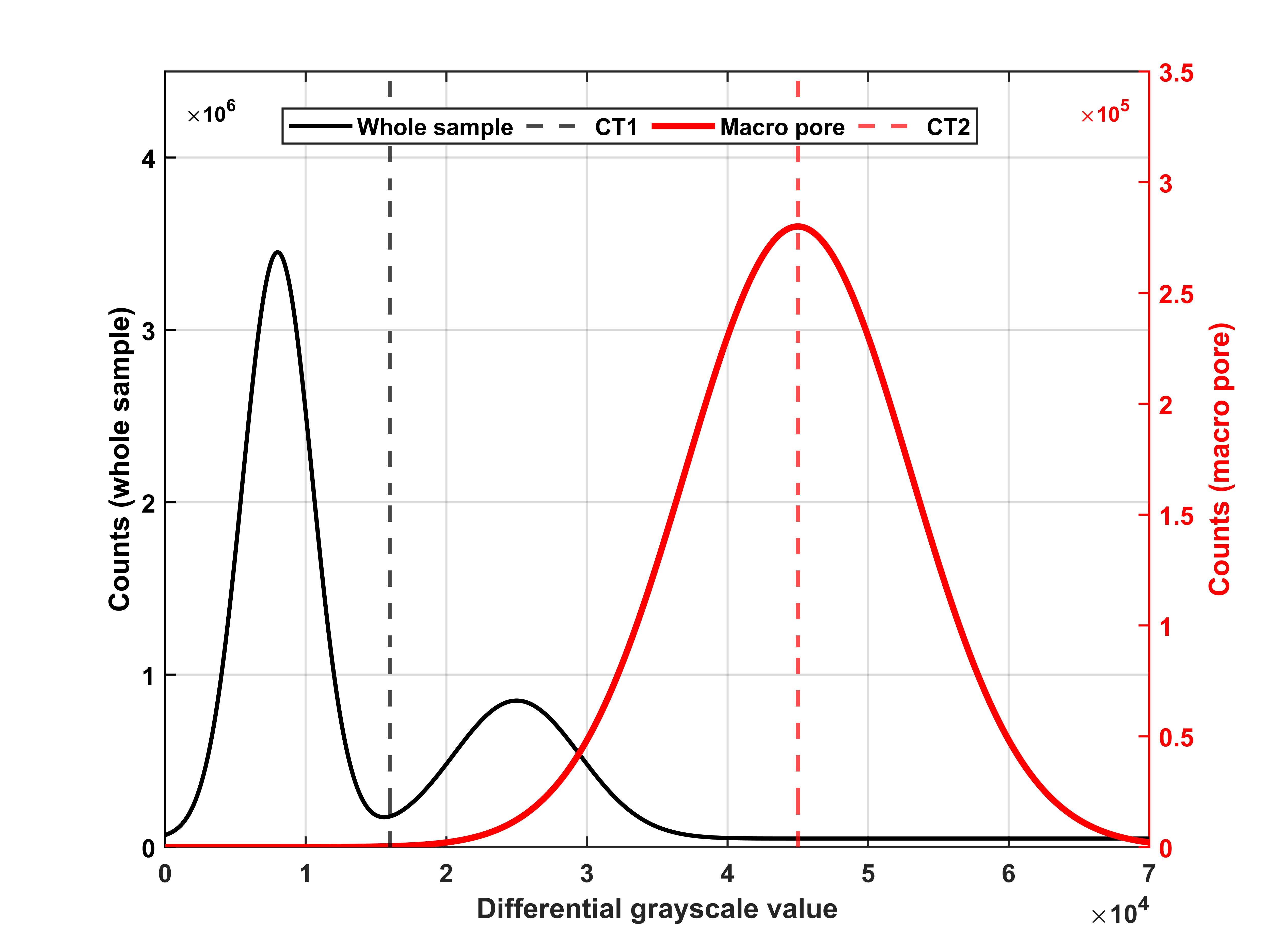}
    \caption{Demonstration of defining thresholds for solid CT1 (0\% porous) and open porosity CT2 (100\% porous), respectively; Note the histogram in this figure comes from synthetic data and does not represent the actual grayscale values of difference images.}
    \label{app_fig:CT1_CT2_Demo_Estaillades}
\end{figure}

The raw images are first processed with the non-local mean filter, then scaled against dry scan with the following equations:

\begin{equation}\label{eq:Normalized_wet_to_dry_scan}
    \begin{aligned}
         I_{new} = (I - p_{s\_brine}) \frac{p_{g\_dry}-p_{s\_dry}}{p_{g\_brine}-p_{s\_brine}}+p_{s\_dry}
    \end{aligned}
\end{equation}

\begin{equation}\label{eq:Normalized_drainage_to_dry_scan}
    \begin{aligned}
         I_{new} = (I - p_{d\_drain}) \frac{p_{g\_dry}-p_{w\_dry}}{p_{g\_drain}-p_{d\_drain}}+p_{w\_dry}
    \end{aligned}
\end{equation}

\noindent where $I_{new}$ is the new scaled image, $I$ is the image before scaling, $p_{s\_brine}$ and $p_{g\_brine}$ are the average of mode grayscale values in three region-of-interest (ROI) of sleeve and grain solid phases in KI saturated scan images, respectively; similarly, $p_{s\_dry}$ and $p_{g\_dry}$ are the average of mode grayscale values in three ROIs of sleeve and grain solid phases in dry scan images, respectively. In equation \ref{eq:Normalized_drainage_to_dry_scan}, $p_{d\_drain}$ is the average of mode grayscale values in three ROIs of decane invaded pores in drainage scans, $p_{g\_drain}$ is the average of mode grayscale values in three ROIs of grain solid phases in drainage scans, and $p_{w\_dry}$ is the average of mode values in three ROIs of water saturated pore regions in dry scan images. After scaling, the difference images of the drainage and the KI saturated scans are calculated as 

\begin{equation}\label{eq:Difference_drain_to_dry_scan}
    \begin{aligned}
         I_{diff\_drain} = I_{new\_drain} - I_{dry}
    \end{aligned}
\end{equation}

\begin{equation}\label{eq:Difference_brine_to_dry_scan}
    \begin{aligned}
         I_{diff\_brine} = I_{new\_brine} - I_{dry}
    \end{aligned}
\end{equation}
The KI saturated-dry scan difference image can be used to calculate the porosity map under the assumption of linear dependence of grayscale value in the difference image and the volume of brine present in each voxel. We define the threshold for $100\%$ solid ($0\%$ porous) and resolved pores voxels as $CT1$ and $CT2$, respectively. $CT1$ is determined by the grayscale value at the valley of the histogram between solid and microporosity phases, see Figure \ref{app_fig:CT1_CT2_Demo_Estaillades}. $CT2$ is determined by masking out the microporosity phases and solid phase regions in the $I_{diff\_brine}$ and finding the mode value of the remaining resolved pore regions, see Figure \ref{app_fig:CT1_CT2_Demo_Estaillades}. Accordingly, the porosity map of microporosity phases $\varphi_{micro}$ is calculated as 

\begin{equation}\label{eq:Porosity_map_calculation}
    \begin{aligned}
         \varphi_{micro} =\frac{I_{diff\_brine}-CT1}{CT2-CT1}
    \end{aligned}
\end{equation}
with porosity map $\varphi_{micro}$ the final whole core porosity is calculated as $0.259$, which is consistent with previous measurements in the literature ($0.247$ to $0.28$).

Accordingly, the saturation map $S_{w\_drain}$ from drainage scans are calculated with equation \ref{eq:Saturation_map_calculation}.

\begin{equation}\label{eq:Saturation_map_calculation}
    \begin{aligned}
         S_{w\_drain} =\frac{I_{diff\_drain}-CT1}{I_{diff\_brine}-CT1}
    \end{aligned}
\end{equation}

\section{Slice saturation comparison of constrained cases}
This section presents the supplementary images for the slice saturation comparison section in the manuscript, which includes local saturation map comparison and slice saturation comparison results of porosity based 3 phases model and K-means based 5 phases model with known connectivity. Here, known connectivity means a 3D mask is created to exclude any non-invaded voxels during each capillary pressure step based on experimental images during regression and intrusion percolation simulation.

\begin{figure}[htp!]
  \centering
    \includegraphics[width=1.0\textwidth]{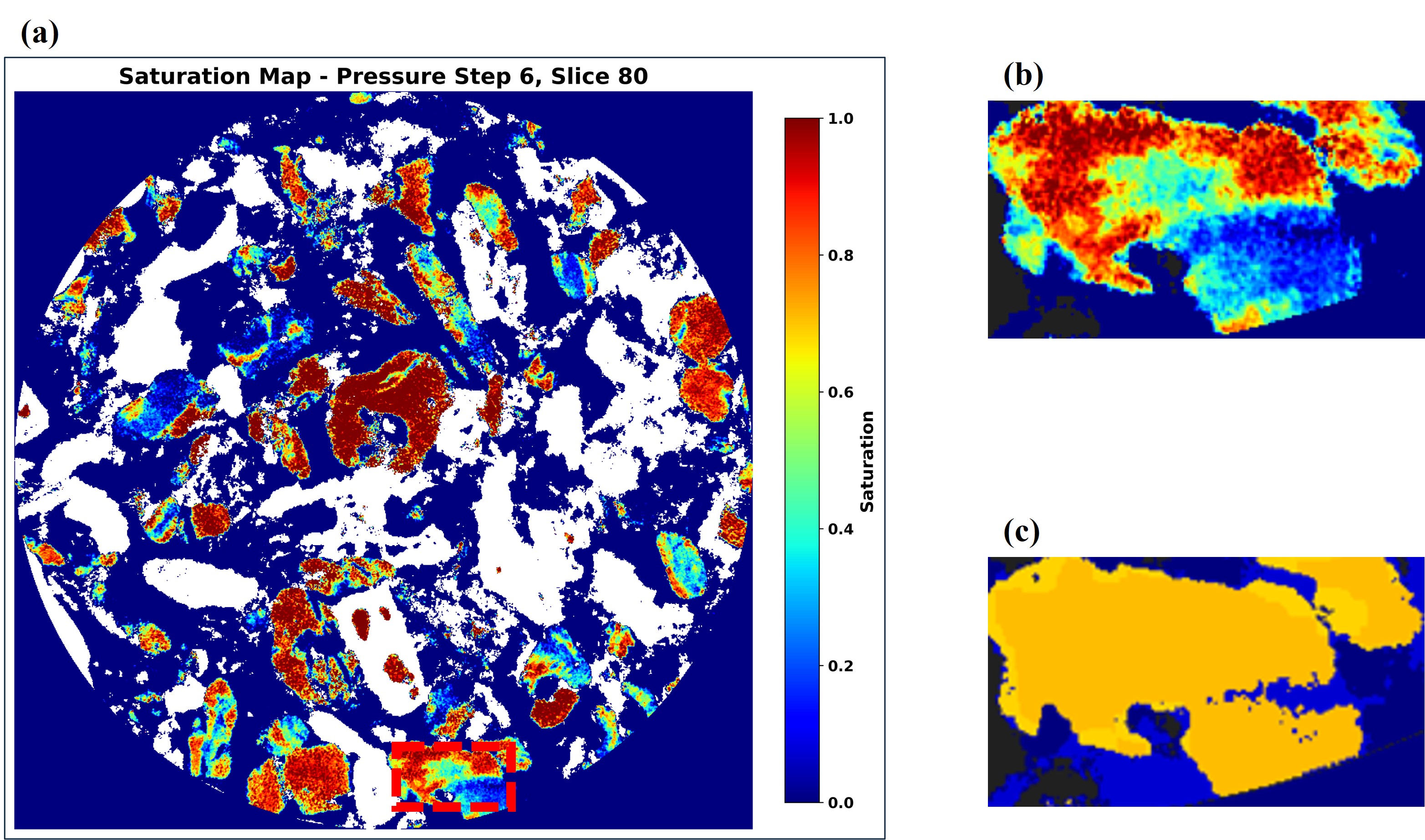}
    \caption{(a) Experimental observation at pressure step 6 slice 80; (b) Local observation of experimental saturation map showing heterogeneous saturation distribution; (c) Porosity-saturation K-means based 5 phases segmentation which smears the local saturation contrast.}
    \label{app_fig:Phases_segmentation_smearing_constrast}
\end{figure}

\begin{figure}[htp!]
  \centering
    \includegraphics[width=1.0\textwidth]{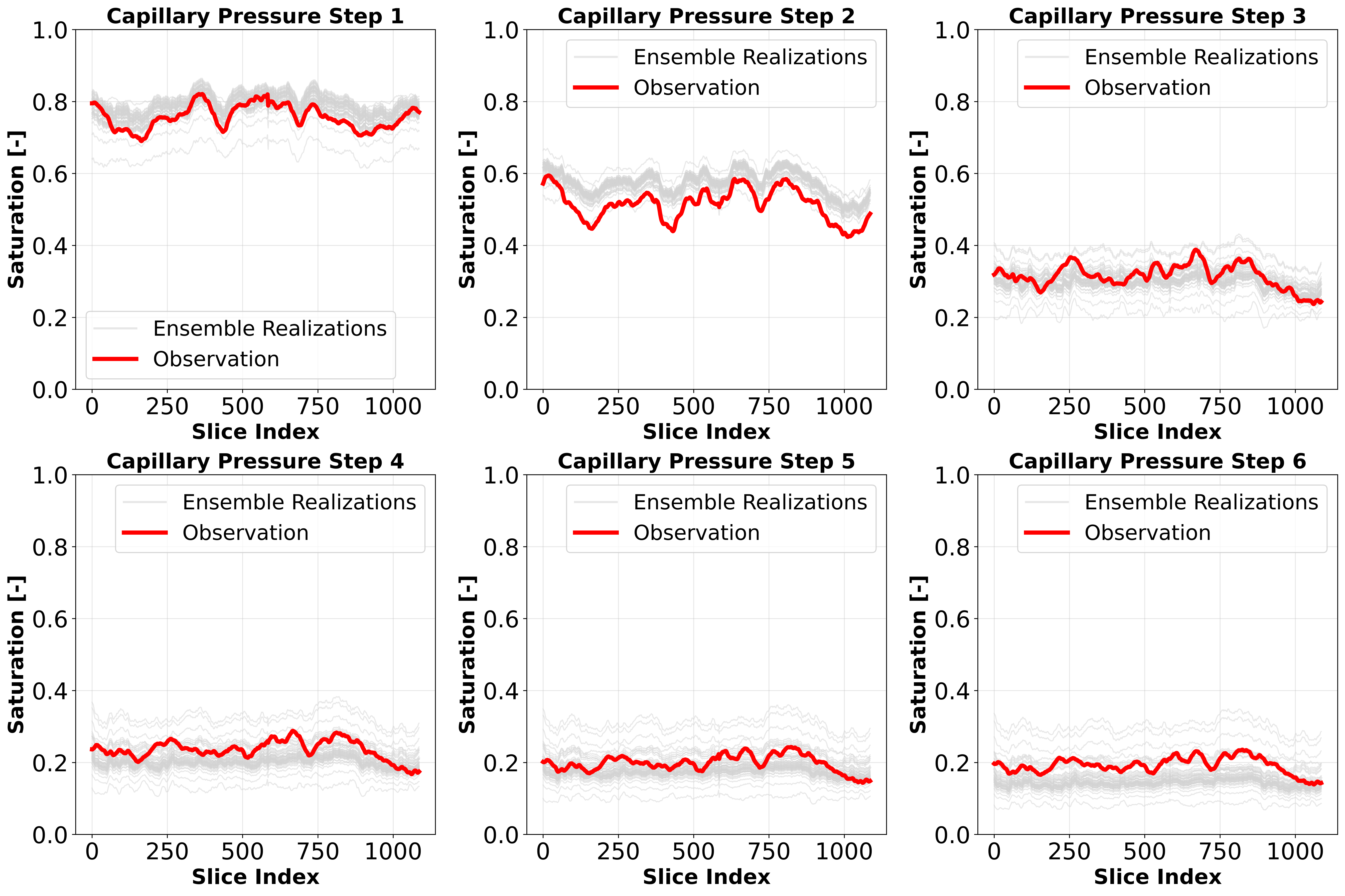}
    \caption{Slice saturation comparison between the intrusion percolation simulation results of the porosity based 3 phases model, with known connectivity, and the experimental images.}
    \label{app_fig:Slice_saturation_comparison_porosity_3_phases_model_connectivity_known}
\end{figure}

%\begin{figure}[htp!]
%  \centering
%    \includegraphics[width=1.0\textwidth]{Figures/Slice_saturation_comparison_K-means_5_phases_model_connectivity_known.png}
%    \caption{Slice saturation comparison between the intrusion percolation simulation results of the porosity-saturation K-means based 5 phases model, with known connectivity, and the experimental images.}
%    \label{app_fig:Slice_saturation_comparison_k-means_5_phases_model_connectivity_known}
%\end{figure}

\section{Porosity-saturation distribution}

\begin{figure}[htp!]
  \centering
    \includegraphics[width=1.0\textwidth]{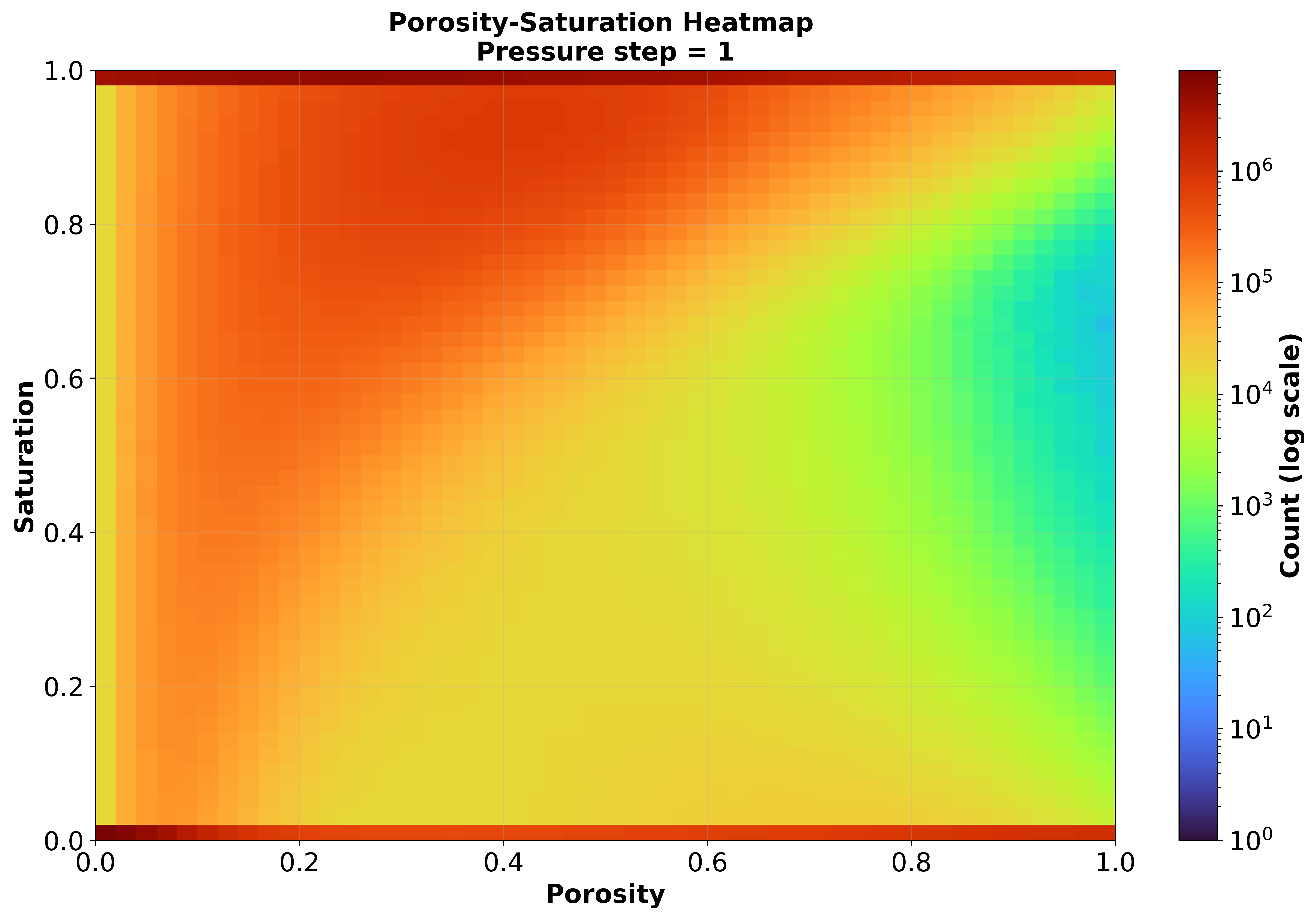}
    \caption{Porosity-saturation distribution heatmap at the first capillary pressure step during the experiment, where high saturation is present in all porosity groups.}
    \label{app_fig: Casestudy_Estaillades_Porosity-saturation_distribution_heatmap}
\end{figure}

\begin{figure}[htp!]
  \centering
    \includegraphics[width=1.0\textwidth]{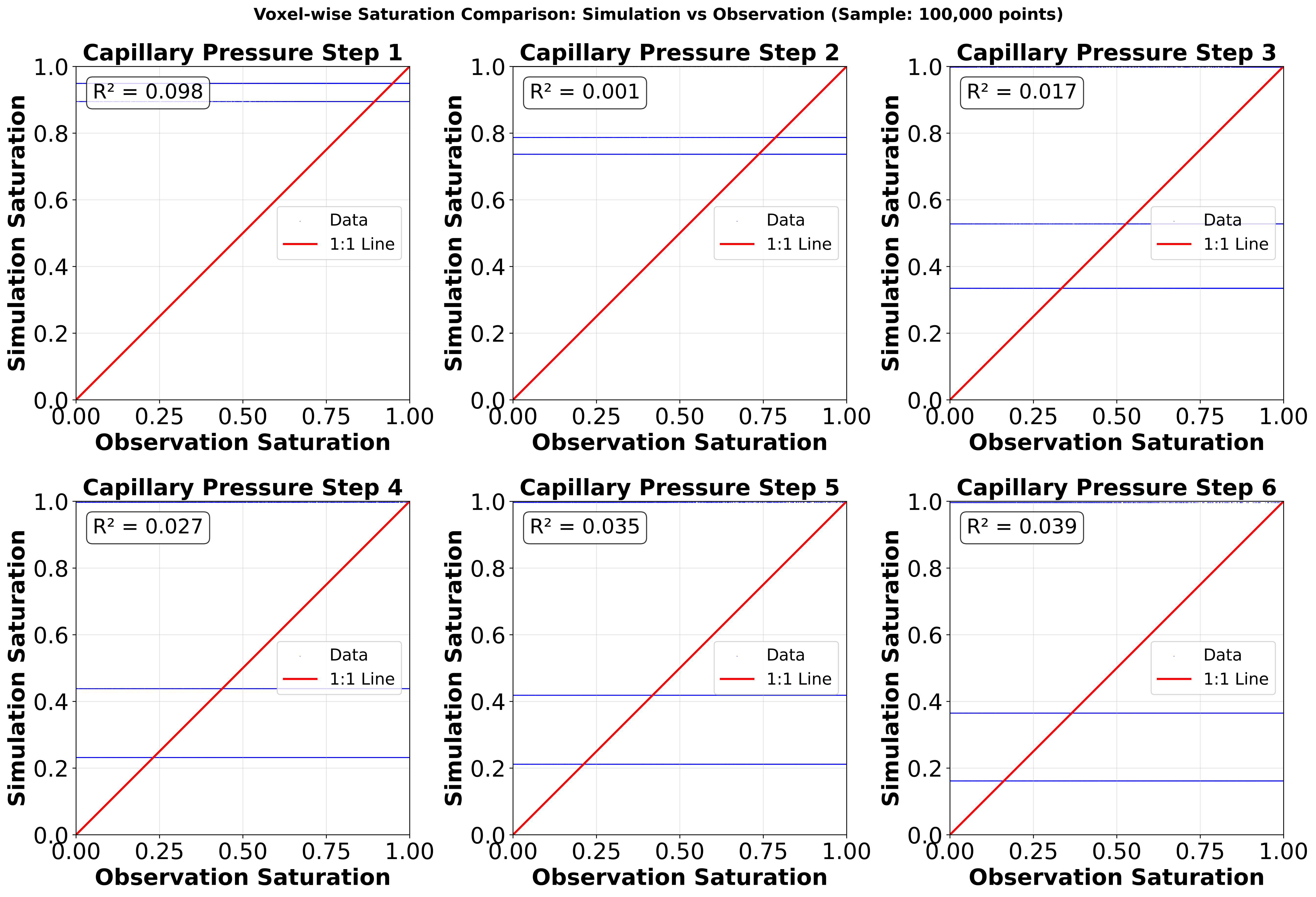}
    \caption{Voxel saturation scatter plot- simulation against experimental observations at each capillary step of porosity based 3 phases model.}
    \label{app_fig:Casestudy_Estaillades_saturation_comparison_scatter_heatmap_3phases}
\end{figure}

\begin{figure}[htp!]
  \centering
    \includegraphics[width=1.0\textwidth]{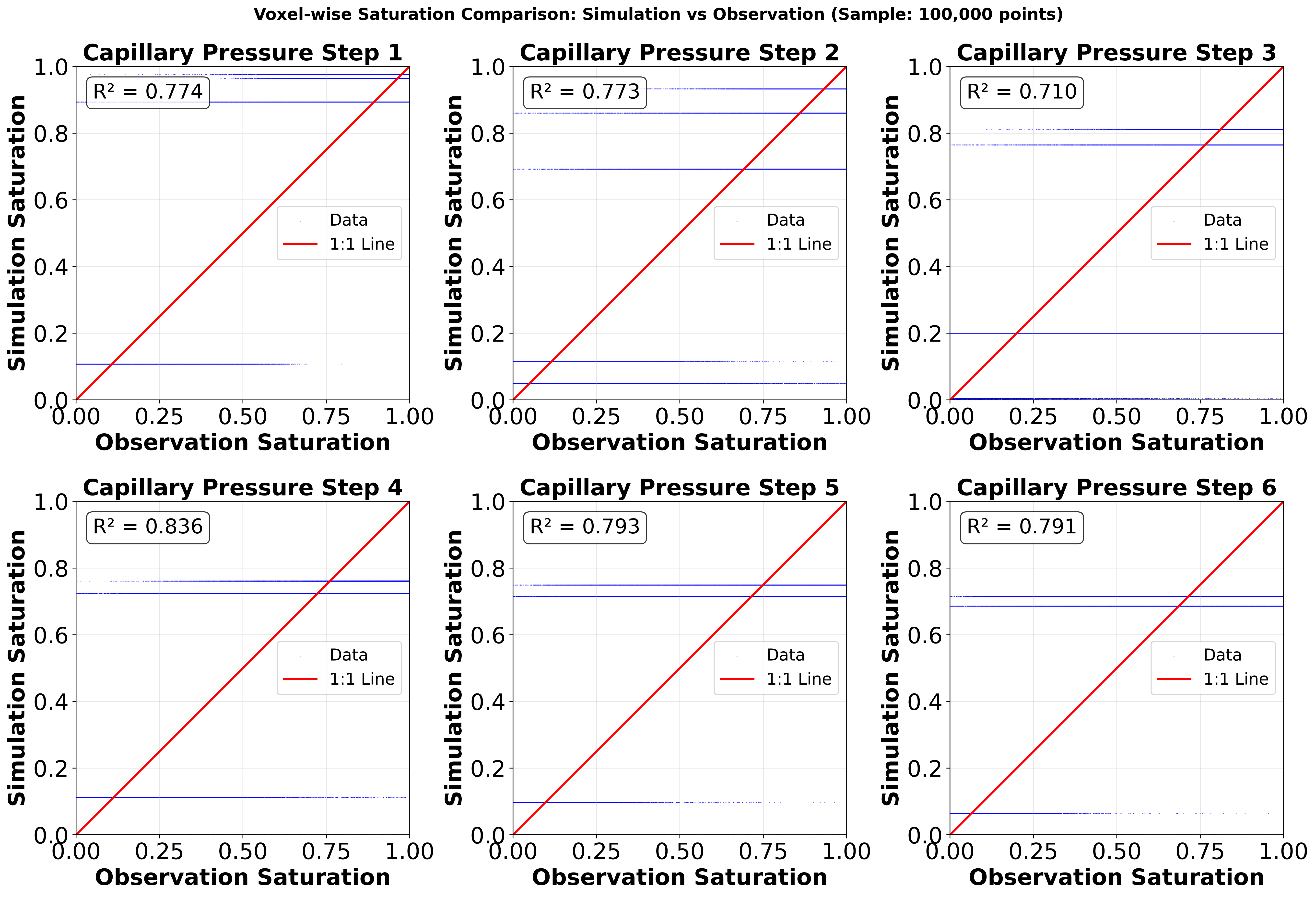}
    \caption{Voxel saturation scatter plot- simulation against experimental observations at each capillary step of porosity-saturation K-means based 5 phases model.}
    \label{app_fig:Casestudy_Estaillades_saturation_comparison_scatter_heatmap_5phases}
\end{figure}

\begin{figure}[htp!]
  \centering
    \includegraphics[width=1.0\textwidth]{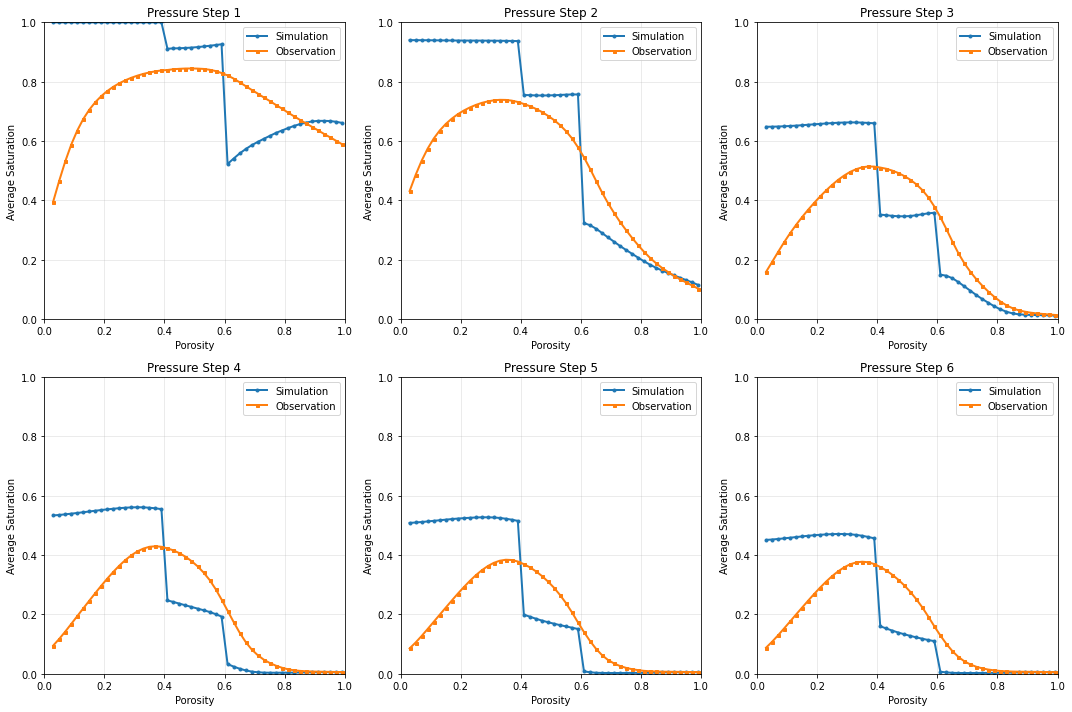}
    \caption{Comparison of average wetting phase saturation distribution across porosity levels between experimental measurements and updated ensembles of unconstrained 3-phase model.}
    \label{app_fig:Casestudy_Estaillades_AvgsSat-por_Distribution_3phases}
\end{figure}

\begin{figure}[htp!]
  \centering
    \includegraphics[width=1.0\textwidth]{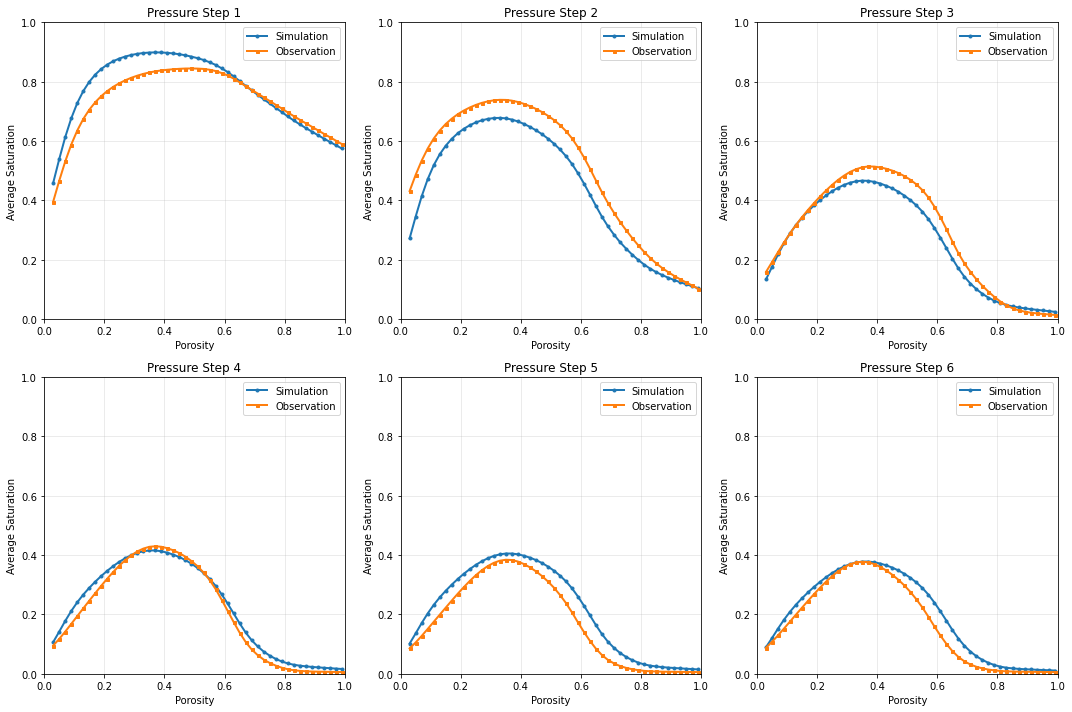}
    \caption{Comparison of average wetting phase saturation distribution across porosity levels between experimental measurements and updated ensembles of unconstrained 5-phase model.}
    \label{app_fig:Casestudy_Estaillades_AvgsSat-por_Distribution_5phases}
\end{figure}

\section{Pore Size Distribution Calculation}\label{app_sec: PSD}

The pore size distribution for the updated ensembles was calculated from the Brooks-Corey capillary pressure model combined with phase distribution data from the 5-phase model.

The Brooks-Corey model relates capillary pressure ($P_c$) to effective saturation ($S_e$) through:
\begin{equation}
P_c = P_e \cdot S_e^{-1/\lambda}
\end{equation}
where $P_e$ is the entry capillary pressure and $\lambda$ is the pore size distribution index.

The Young-Laplace equation converts capillary pressure to pore throat radius ($r$):
\begin{equation}\label{app_eq:Yong_equation}
P_c = \frac{2\sigma \cos(\theta)}{r}
\end{equation}
where $\sigma = 0.030$ N/m is the oil-brine interfacial tension and $\theta = 30^o$ is the contact angle (water-wet condition).

The pore size distribution was calculated through the following steps:

\begin{enumerate}
    \item \textbf{Phase filtering}: Voxels within the target porosity range (0.2--0.5) were extracted from the porosity map, yielding phase distributions and voxel counts for five distinct phases.
    
    \item \textbf{Individual phase PSDs}: For each phase $i$ with parameters $(P_{e\_i}, \lambda_i)$, the capillary pressure curve was calculated over the saturation range $S_e \in [0.001, 0.999]$, then converted to pore throat radius using Eq. \ref{app_eq:Yong_equation}.
    
    \item \textbf{Frequency calculation}: The pore size frequency distribution for each phase was computed as:
    \begin{equation}
    f_i(r) = \left| \frac{dS_e}{dr} \right|
    \end{equation}
    
    \item \textbf{Overall distribution}: The overall pore size frequency distribution was obtained by weighting each phase by its volumetric fraction ($w_i$):
    \begin{equation}
    f_{overall}(r) = \sum_{i=1}^{5} w_i \cdot f_i(r)
    \end{equation}
    where $w_i = N_i / N_{total}$, with $N_i$ being the number of voxels in phase $i$.
    
    \item \textbf{Cumulative integration}: The cumulative distribution was calculated by integrating the overall pore size frequency distribution, and the final frequency plot represents the percentage changes in cumulative pore volume at each pore size.
    
    \item \textbf{Physical cutoff}: A minimum pore size of 0.001 \textmu m was applied to exclude sub-nanometer scales that are not physically representative of rock pores.
\end{enumerate}

\section{Acknowledgments}
This work is funded by the Engineering and Physical Sciences Research Council's ECO-AI Project grant (reference number EP/Y006143/1), with additional financial support from the PETRONAS Centre of Excellence in Subsurface Engineering and Energy Transition (PACESET).

\section{Data availability}
The micro-CT images of the Estaillades rock sample are available via \cite{wang2022anchoring}. The implementation code of the ESMDA algorithm is available at \url{https://github.com/DigiPorFlow/Estaillades_BC_ESMDA}.  

\bibliographystyle{elsarticle-harv}
\bibliography{Reference.bib}

\end{document}